\documentclass[twocolumn,times]{aastex701}
\usepackage{amsmath}
\usepackage{longtable}
\usepackage{float}
\usepackage{mathrsfs}
\usepackage{gensymb}
\usepackage{mathtools}

\begin{document}

\title{TIME Commissioning Observations: II. On-sky Characterization and the 2D Map Data Processing Pipeline}

\author[0000-0002-9813-0270]{Benjamin J.~Vaughan}
\affiliation{Department of Physics, Cornell University, Ithaca NY 14853, USA}
\email{bjv37@cornell.edu}

\author[0009-0005-4099-8842]{Abigail T.~Crites}
\affiliation{Department of Physics, Cornell University, Ithaca NY 14853, USA}
\affiliation{Department of Astronomy, Cornell University, Ithaca NY 14853, USA}
\affiliation{Department of Physics, California Institute of Technology, Pasadena, California 91125, USA}
\email{atc72@cornell.edu}

\author[0000-0003-2618-6504]{Dongwoo T.~Chung}
\affiliation{Department of Astronomy, Cornell University, Ithaca NY 14853, USA}
\email{dongwooc@cornell.edu}

\author{Ryan P.~Keenan}
\affiliation{Max-Planck-Institut f\"{u}r Astronomie, K\"{o}nigstuhl 17, D-69117 Heidelberg, Germany}
\email{keenan@mpia.de}

\author{James~J.~Bock}
\affiliation{Department of Physics, California Institute of Technology, Pasadena, California 91125, USA}
\affiliation{Jet Propulsion Laboratory, California Institute of Technology, Pasadena, California 91109, USA}
\email{jjb@astro.caltech.edu}
\author{Charles~M.~Bradford}
\affiliation{Department of Physics, California Institute of Technology, Pasadena, California 91125, USA}
\affiliation{Jet Propulsion Laboratory, California Institute of Technology, Pasadena, California 91109, USA}
\email{bradford@submm.caltech.edu}
\author[0000-0002-0941-0407]{Victoria L.~Butler}
\affiliation{Department of Physics, Cornell University, Ithaca NY 14853, USA}
\email{vlb59@cornell.edu}

\author{Tzu-Ching~Chang}
\affiliation{Jet Propulsion Laboratory, California Institute of Technology, Pasadena, California 91109, USA}
\email{tzu-ching.chang@jpl.nasa.gov}
\author{Yun-Ting~Cheng}
\affiliation{Department of Physics, California Institute of Technology, Pasadena, California 91125, USA}
\affiliation{Jet Propulsion Laboratory, California Institute of Technology, Pasadena, California 91109, USA}
\email{ycheng3@caltech.edu}
\author[0009-0007-6638-5774]{Audrey~Dunn}
\affiliation{Rochester Institute of Technology, Rochester, NY, 14623, USA}
\email{akd5648@g.rit.edu}
\author{Nicholas~Emerson}
\affiliation{Department of Astronomy and Steward Observatory, University of Arizona, 933 N Cherry Avenue, Tucson, AZ 85721, USA}
\email{nemerson@arizona.edu}
\author{Clifford~Frez}
\affiliation{Jet Propulsion Laboratory, California Institute of Technology, Pasadena, California 91109, USA}
\email{Clifford.F.Frez@jpl.nasa.gov}
\author{Jonathon~Hunacek}
\affiliation{Jet Propulsion Laboratory, California Institute of Technology, Pasadena, California 91109, USA}

\email{jhunacek@gmail.com}
\author{Chao-Te~Li}
\affiliation{Institute of Astronomy and Astrophysics, Academia Sinica, Taipei, Taiwan}
\email{ctli@asiaa.sinica.edu.tw}
\author[0000-0003-4063-2646]{Ian~N.~Lowe}
\affiliation{Department of Astronomy and Steward Observatory, University of Arizona, 933 N Cherry Avenue, Tucson, AZ 85721, USA}
\email{ianlowe@arizona.edu}

\author{King~Lau}\affiliation{Department of Physics, California Institute of Technology, Pasadena, California 91125, USA}
\email{kennylau@caltech.edu}
\author{Daniel~P.~Marrone}
\affiliation{Department of Astronomy and Steward Observatory, University of Arizona, 933 N Cherry Avenue, Tucson, AZ 85721, USA}
\email{dmarrone@email.arizona.edu}
\author[0000-0001-6439-8140]{Evan~C.~Mayer}
\affiliation{Department of Astronomy and Steward Observatory, University of Arizona, 933 N Cherry Avenue, Tucson, AZ 85721, USA}
\email{evanmayer@arizona.edu}

\author[0009-0002-1152-4953]{Sophie M.~McAtee}
\affiliation{Department of Physics, Cornell University, Ithaca NY 14853, USA}
\email{smm549@cornell.edu}

\author{Dang Pham}
\affiliation{JILA and Department of Astrophysical and Planetary Sciences, CU Boulder, Boulder, CO 80309, USA}
\email{dang.pham@astro.utoronto.ca}
\affiliation{David A.~Dunlap Department of Astronomy and Astrophysics, University of Toronto, 50 St. George Street, Toronto, ON M5S 3H4, Canada}

\author[0009-0004-1047-3714]{Shwetha Prakash}
\affiliation{Department of Physics, Cornell University, Ithaca NY 14853, USA}
\email{sp2358@cornell.edu}

\author{Guochao~Sun}
\affiliation{CIERA and Department of Physics and Astronomy, Northwestern University, 1800 Sherman Avenue, Evanston, IL 60201, USA}
\email{jsun.astro@gmail.com}
\author{Isaac~Trumper}
\affiliation{Department of Astronomy and Steward Observatory, University of Arizona, 933 N Cherry Avenue, Tucson, AZ 85721, USA}
\email{itrumper@optics.arizona.edu}
\author{Anthony~D.~Turner}
\affiliation{Jet Propulsion Laboratory, California Institute of Technology, Pasadena, California 91109, USA}
\email{anthony.d.turner@jpl.nasa.gov}
\author{Ta-Shun~Wei}
\affiliation{Institute of Astronomy and Astrophysics, Academia Sinica, Taipei, Taiwan}
\email{tashun@asiaa.sinica.edu.tw}

\author[0009-0002-7777-2351]{Selina F. Yang}
\affiliation{Department of Physics, Cornell University, Ithaca NY 14853, USA}
\email{fy235@cornell.edu}

\author[0000-0001-8253-1451]{Michael~Zemcov}
\affiliation{Rochester Institute of Technology, Rochester, NY, 14623, USA}
\affiliation{Jet Propulsion Laboratory, California Institute of Technology, Pasadena, California 91109, USA}
\email{zemcov@cfd.rit.edu}

\collaboration{all}{The TIME Collaboration}

\begin{abstract}
    The Tomographic Ionized-carbon Mapping Experiment (TIME) is a line intensity mapping (LIM) instrument that is designed to observe the power spectrum of the [CII] $158$~$\mu$m emission line during the Epoch of Reionization. TIME completed a commissioning run in 2022 at the Arizona Radio Observatory 12-meter Radio Telescope, on Kitt Peak, where it observed Galactic sources for the first time. In this paper we report on an analysis of observations of the Orion Molecular Cloud (OMC) and G49.5 (a local HII region). The OMC observations were taken at least once a day to assess the stability of the instrument and demonstrate its on-sky performance. We describe a spectral image processing pipeline to make calibrated maps of raster scans of these sources, incorporating planet observations for gain calibration. We show with G49.5 that, when compared to the Bolocam Galactic Plane Survey, we are able to achieve a $< 2\%$ calibration difference. Based on the outcomes from this commissioning phase of TIME, we have demonstrated preliminary performance, and identified sources of improvement necessary for pursuing a LIM measurement. 

\end{abstract}

\section{Introduction}

We have deployed the Tomographic Ionized-carbon Mapping Experiment (TIME) at the Arizona Radio Observatory 12 Meter (ARO-12m) telescope and in this paper present on-sky characterization measurements during the 2022 commissioning. TIME \citep{TIME_pilot} is an imaging spectrometer that operates in a frequency range of $183$-$326$~GHz with a spectral resolution of $R\sim100$. TIME's primary science cases are: observations of the [CII] $158$~$\mu$m forbidden transition line emitted during the epoch of reionization over redshifts $6 \lesssim z \lesssim 9$; and observations of CO rotational lines during cosmic noon ($1 \lesssim z \lesssim2$). These measurements allow us to probe the history of star formation rate during these two epochs; The [CII] measurements provide direct access to the luminosity of photodisassociation regions, and CO measurements give an estimation of the molecular hydrogen gas content \citep{Sun2021,kovetz_2019}. With TIME, we access this information by producing a set of tomographic intensity maps, and computing the 3-D power spectrum of these emission lines. The design requirements needed to achieve this goal, low spectral resolution over a wide spectral range, also permit the production of a wealth of information about molecular clouds, planetary nebulae, and other forms of spatially extended mm-wavelength emission. 

To prepare for LIM measurements, TIME underwent a commissioning season culminating in $11$ days of on-sky observations during the winter of 2022. A variety of objects were studied including local molecular clouds (this work), the Galactic center \citep{Yang2026}, planets, and quasars. These observations are split into two categories: 1-D line scans largely designed to characterize the noise in the instrument, and 2-D rasters. Here we focus on the 2-D raster observations of planets and molecular clouds that allow us to characterize the full on-sky performance of TIME. In particular, we produce calibrated spectral images of the Orion Molecular Cloud (OMC) complex \citep{Bally_2008} which was observed at least once a day throughout the whole commissioning, and G49.5, a local HII region \citep{Carpenter_1998}. The OMC complex and G49.5 are two of the brightest radio sources in the sky with significant CO (2-1) emission \citep{Fujita_2021,Wilson_2005}, making them well-studied and suited for understanding the performance of TIME. 

This paper is structured as follows. In \S\ref{sec:state_of_inst} we describe the status of the TIME focal plane, the observing strategy and how the raw data are packaged. In \S\ref{sec:map_making}, the process of converting these raw data to on-sky maps and calibrating observations to physical units is described. In \S\ref{sec:g49}, we process one G49.5 observation and compare its flux against the Bolocam Galactic Plane Survey (BGPS, \citet{Ginsburg2013}). In \S\ref{sec:OMC}, we process observations of the Orion Molecular Cloud complex (OMC), and in \S\ref{sec:discussion}, we discuss the implications of these results for future TIME surveys.  

\section{Instrument and Observations}
\label{sec:state_of_inst}

\begin{figure*}[]
    \centering
     \includegraphics[width=1\linewidth]{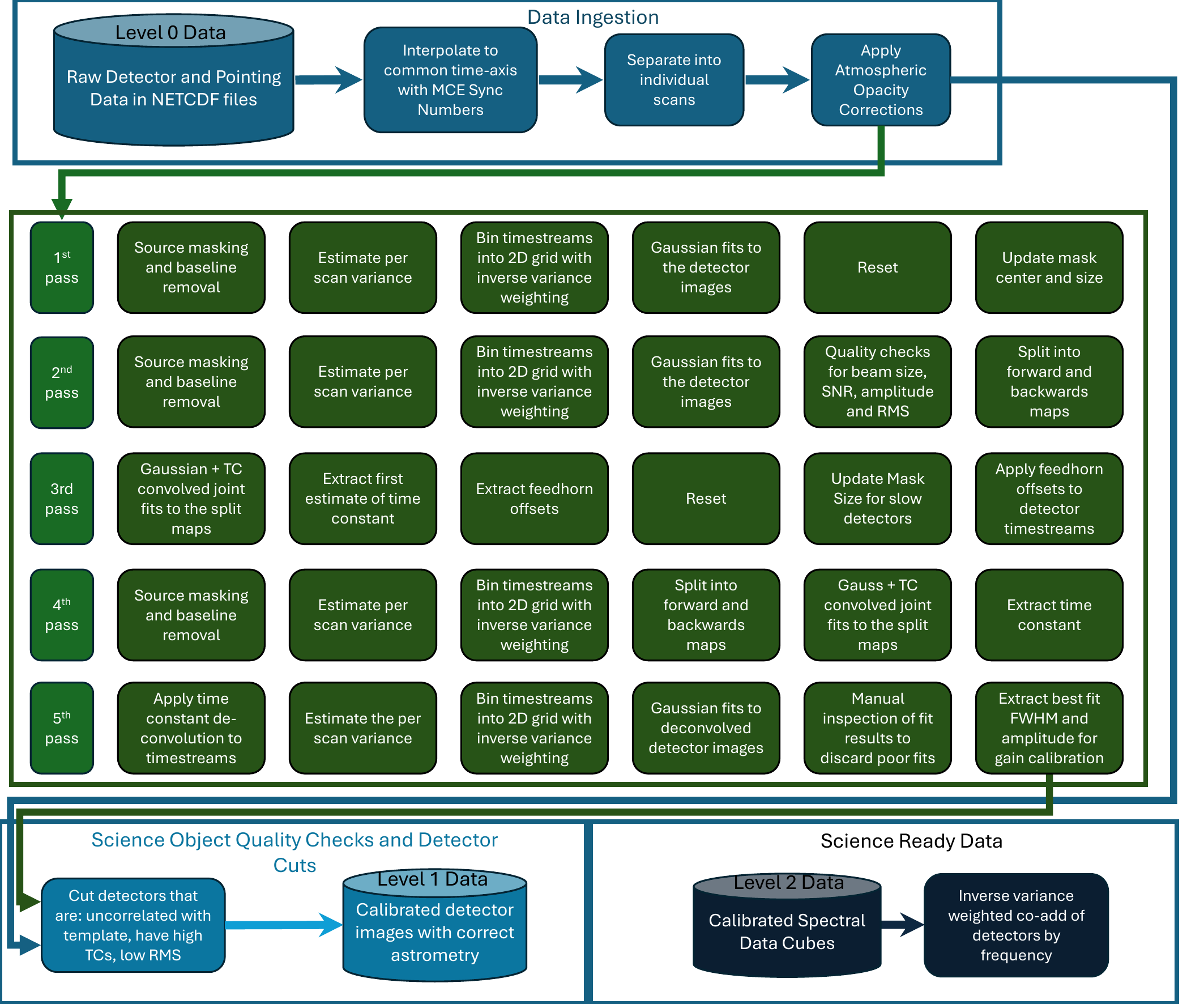}
    \caption{A flow chart describing the map-making process. We start with a planet observation; it is first processed using a peak finding algorithm to mask the planet signal during baseline detrending. Next, Gaussian beams are fit to these data and the quality control checks are applied. The data are then split into sets of forward-scan and backward-scan maps. These are then fit to using a joint fit across both maps that accounts for the time constants and group delay described in equations \ref{eq:deconv_kernel} and \ref{eq:transfer_func}. At this point the relative feedhorn offsets are extracted and applied to the later steps. We also update the mask baseline subtraction mask by multiplying it by a scaling factor proportional to the time constant where slower time constants get a higher multiple. We then recompute the forward and backward maps using this new mask and re-fit the joint time constant model. The time constants from the fourth pass fit are then extracted and used to inform the deconvolution kernel. Each detector is then deconvolved in timestream space and a new set of maps is computed using both scan directions. We then fit a Gaussian beam to the data and extract the gains following the methodology described in equation \ref{eq:gain}.}
    \label{fig:flowchart}
\end{figure*}

TIME is comprised of two grating spectrometer banks that each have sixteen feedhorns arranged in a line \citep{Chao-te_2018}. Light that enters the TIME cryostat is split by a wire grid into two orthogonally polarized beams, each of which feeds a separate waveguide spectrometer. The detectors are transition edge sensor (TES) bolometers \citep{TIME_TES, Hunacek_2018, SPIE_proceeding} and are time-domain multiplexed via a multichannel electronics (MCE; \citealt{Battiselli_2008}) board. During observations, the TIME cryostat is mounted in the receiver cabin of the ARO-12m telescope. TIME is then optically coupled to the telescope through a set of six aluminum mirrors \citep{Mayer_2026, Trumper_2019}. The last three of these mirrors form a K-mirror configuration to correct the parallactic angle rotation induced by the alt-az mounting of the telescope. 

\subsection{Status of the TIME Focal Plane}

For the 2022 commissioning run, TIME had just one of two spectrometer banks installed and therefore operated as a single-polarization instrument. Each spectrometer is populated with 
$24$ detector modules (see \citet{SPIE_proceeding} for a full description), which consist of two physically different varieties, low frequency ($183$-$230$~GHz) and high frequency ($230$-$323$~GHz). These modules are then mounted onto the back-side of the diffraction gratings \citep{Chao-te_2018}; during this run, unpopulated modules were used to block stray light from portions of the focal plane with missing detectors. We had a total of $200$ to $300$ fully functioning detectors and were able to recover $58$ out of the $60$ spectral channels at $R\sim100$.  

\subsection{Observing Strategy and Raw Data}
\label{sec:data}

Over the $11$-day 2022 commissioning run, TIME made over $300$ observations across a total of $\sim 203$ hours. Throughout this session, the observing strategy was iterated and improved. We discuss now the generic observing strategy for a TIME observation, which is a simple 2-D raster for extended radio sources. In this scheme, we hold one coordinate constant (e.g. declination) and scan the instrument across the other (e.g. right ascension). We then step to the next constant-coordinate grid point, scan in the ``fast'' coordinate, and repeat this process until we have mapped out a 2-D grid on the sky.

There are a few factors to keep in mind for optimizing scan parameters. One example is the impact of scan traversal time on the baseline drifts we see from $1/f$ noise present in the data. Another is the detector physics for the TES bolometers used in TIME. The detector time constants vary and can be comparable to the beam transit time for reasonable scan speeds, resulting in an apparent smearing of the beam in the scan direction (e.g., \citet{Jon_PHD} chapter 5.2); their effect is shown in Figure \ref{fig:deconvolution_showcase}, and we have built a data processing pipeline that can model them.

\begin{deluxetable*}{lcccccc}
    \tablecaption{Observing Parameters}
    \label{tab:obs_table}
    \tablehead{\colhead{Object} & \colhead{$N_{\rm obs}$} & \colhead{$\langle \Delta {\rm El} \rangle$ [deg]} & \colhead{\shortstack{Vertical \\ Raster Step [arcsec]}} & \colhead{Scan Speed [deg s$^{-1}$]} & \colhead{RA (J2000)} & \colhead{Dec (J2000)}}
    \startdata
    OMC-1 A & $2$ & $10.6$ & $3.6$ & $0.0625$ & 05:35:09.8 & -05:24:58 \\
    OMC-1 B & $3$ & $5.2$ & $3.6$ & $0.1$ & 05:35:12.9 & -05:22:26 \\
    OMC-1 C & $7$ & $4.3$ & $18$ & $0.1$ & 05:35:18.3 & -05:22:31 \\
    G49.5 A & $1$ & $4.1$ & $3.6$ & $0.1$ & 19:23:43.2 & +14:30:36 \\
    Jupiter A & $1$ & $1.5$ & $3.6$ & $0.1$ & Planet & Planet \\
    Jupiter B & $2$ & $0.4$ & $7.2$ & $0.1$ & Planet & Planet \\
    Jupiter C & $1$ & $1.8$ & $18$ & $0.075$ & Planet & Planet \\
    Venus A & $1$ & $0.6$ & $7.2$ & $0.05$ & Planet & Planet \\
    Mars A & $1$ & $0.4$ & $3.6$ & $0.0375$ & Planet & Planet \\
    Mars B & $1$ & $3.1$ & $3.6$ & $0.15$ & Planet & Planet \\
    Mars C & $1$ & $1.2$ & $3.6$ & $0.075$ & Planet & Planet \\
    \enddata
    \tablecomments{Observations grouped by object and scan parameters: a group shares one vertical raster step and one scan speed. The elevation column gives the range spanned by the on-source data of every observation in the group. Coordinates are the median pointing center.}
\end{deluxetable*}

\subsection{Atmospheric Effects}
 The ARO-12m normally monitors the 225 GHz atmospheric opacity, $\tau_{225}$, using a tipping radiometer. However, the tipper was not operating during the 2022 TIME commissioning run. 
 In this work, we instead derive opacity estimates using an atmospheric modeling code and contemporary forecasts from the NOAA Global Forecast System (GFS). GFS predictions are released with 28~km spatial resolution every 6 hours, and the forecasted atmosphere is translated to zenith opacity with the 
 \textit{am} atmospheric modeling software \citep{Paine2019}. 
 We extrapolate $\tau_{225}$ to $\tau_\nu$ for the nominal frequency of each detector again using \textit{am}, assuming a standard mid-latitude atmosphere and an elevation of $1900$~m.

Since the opacity at Kitt Peak is relatively high ($\tau_{225} \sim 0.1$ at zenith in the top 10\% of conditions during the 2022 observing season), the atmosphere produces systematics that need to be accounted for. The atmospheric effects can be divided into two primary  components: atmospheric attenuation, which weakens astrophysical signals; and atmospheric emission, which generates a time-varying source of photons with a $1/f$ spectrum whose characteristics are dependent on meteorological conditions and observed frequency \citep{Sayers_2010}.

\subsection{OMC and G49.5 Observations}

The raw data for TIME observations comes in the form of time-ordered-data (TOD) for the telescope's pointing, operational flags (e.g. whether or not the telescope is in a turnaround zone), and each detector's flux measurement. The telescope signals are then interpolated onto the detector timestreams. We use the operational data to sort the pointing and detector TODs into individual raster scans and process those into maps using a software pipeline that we call \textit{SPACETIME} (Sky Pointing and Amplitude Calibration Estimator for TIME). This pipeline implements the procedure outlined in Figure \ref{fig:flowchart} and the following section.

\section{Map Making and Calibration}
\label{sec:map_making}
A critical step towards analyzing these data is their projection into an image. As described in the preceding section, and outlined in Figure \ref{fig:flowchart}, we start with a set of TOD that has been sorted by raster scan. Then we correct for atmospheric attenuation;
\begin{equation}
   S = S_0e^{-\tau(\nu)\sec(z)}.
\end{equation}
Above, $\tau(\nu)$ is the atmospheric opacity as a function of frequency, $z$ is the zenith angle, $S$ is the measured flux density and $S_0$ is the flux density at the source. Then we mask the source and subtract a fit to a 1st order polynomial to remove baseline drifts in the detector timestreams, caused in part by time-varying atmospheric emission and drifts in the stage temperature. For bright and localized sources such as planets and quasars we start the map-making process by utilizing a peak finding algorithm to mask the source in the timestreams; for other sources we produce a custom mask. After baseline corrections we compute the per-scan variance from the off source components of the timestreams. Finally, we project these timestreams to maps with a 2-D histogram where we estimate the flux per pixel with an inverse variance weighting of its constituents. We then fit beams to these data, update the masks and extract calibration products. Since each frequency channel has a different spectral response function and each detector has its own gain, this is done on a detector by detector basis. 
\subsection{Planet Calibration}
\label{sec:planet_cal}

\begin{figure*}[ht!]
    \centering
    \includegraphics[width=1\linewidth]{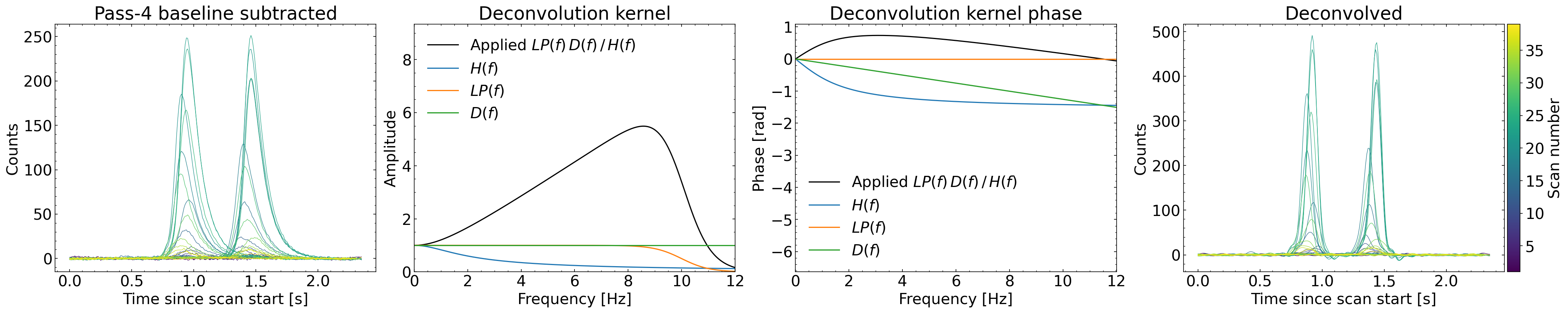}
    \caption{A schematic of the deconvolution that takes place for a set of timestreams. In the left panel is a set of timestreams that have had their baseline subtracted. There are two peaks corresponding to forward and backwards maps that are offset since the start time of the scans. In the middle is the deconvolution kernel that is used. We show the transfer function of the detector's time constant and include a low pass filter to limit boosting of high frequency noise. On the right is the set of timestreams after they have been deconvolved where the long tail has been suppressed. }
    \label{fig:scan_deconvolution}
\end{figure*}

The $16$ feedhorn layout of the focal plane means that there are $16$ separate lines of sight at any given time in an observation that produce a $\sim 13.5\arcmin \times 0.5\arcmin$ linear beam pattern. We describe here the process of aligning these different views with the operational pointing data from the ARO-12m by estimating the feedhorn pointing offsets.

The process to assess these planet calibration products is done in a five step iterative process, as outlined in Figure \ref{fig:flowchart}. The first step, which was described in the previous section, is to fit a beam to a planet map generated via the naive scan filtering. From there, we can use the best fit centers of those data to re-estimate where to place the per-scan mask, and then recompute the baseline drifts accordingly. In this second pass we are able to mitigate effects of over-subtraction seen in some detector data. We then re-fit a planet disk convolved with a beam model to these data and perform a set of four quality checks to remove detectors lacking signal. Two of these checks are to reduce the effect of cross-talk and are handled by computing the SNR of the best fit amplitude as well as a minimum required amplitude of detector counts. The two other checks are to see if there are detectors with only white noise signals present in the data. Since a reduced chi squared of their best fit would still produce a value of one, the ratio of the RMS of the detector map before and after subtracting the best fit model is used to cull poorly fit detectors, in combination with a check that the FWHM is above $\lambda/D$, where D is the size of the telescope dish. These four checks ensure that the fitted beam models are physically meaningful. 

\begin{figure}[h!]
    \centering
    \includegraphics[width=1\linewidth]{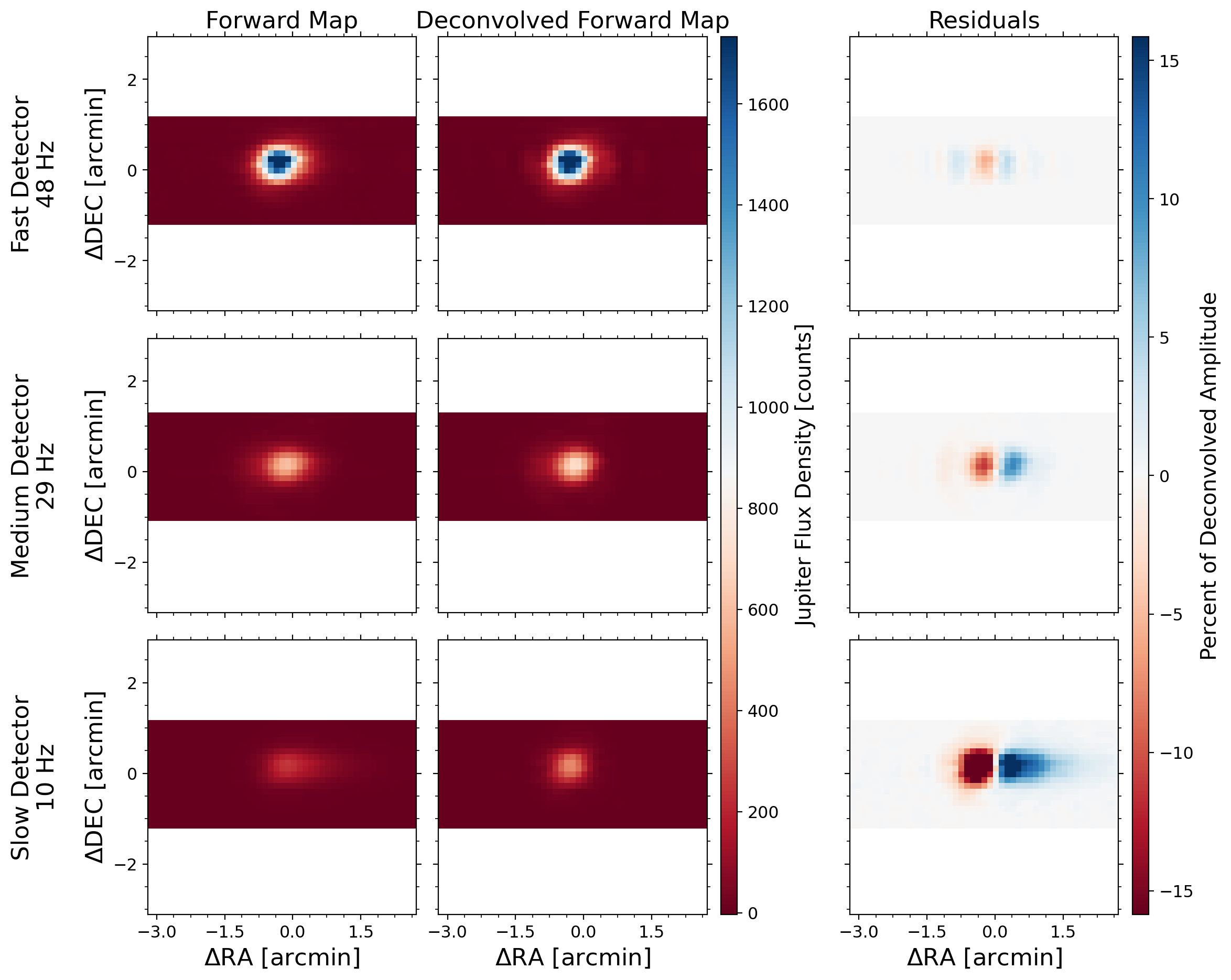}
    \caption{Shown in the top 9 panels are the effect of this deconvolution procedure on a fast detector, a medium speed detector, and a slow detector. In the fast detector there is still some residual difference which is due to the difference in group delay from readout electronics that becomes more apparent once the time constant effects are modeled with a deconvolution, this is also shown in the middle portion. In the slow detector we see that there is a very clear tail in the pre-deconvolved data that gets accurately modeled with the deconvolution method. The bottom 4 panels show a set of example group delay tests where we show some example peaks compared and offset as a function of group delay. }
    \label{fig:deconvolution_showcase}
\end{figure}

\begin{figure}
    \centering
    \includegraphics[width=1\linewidth]{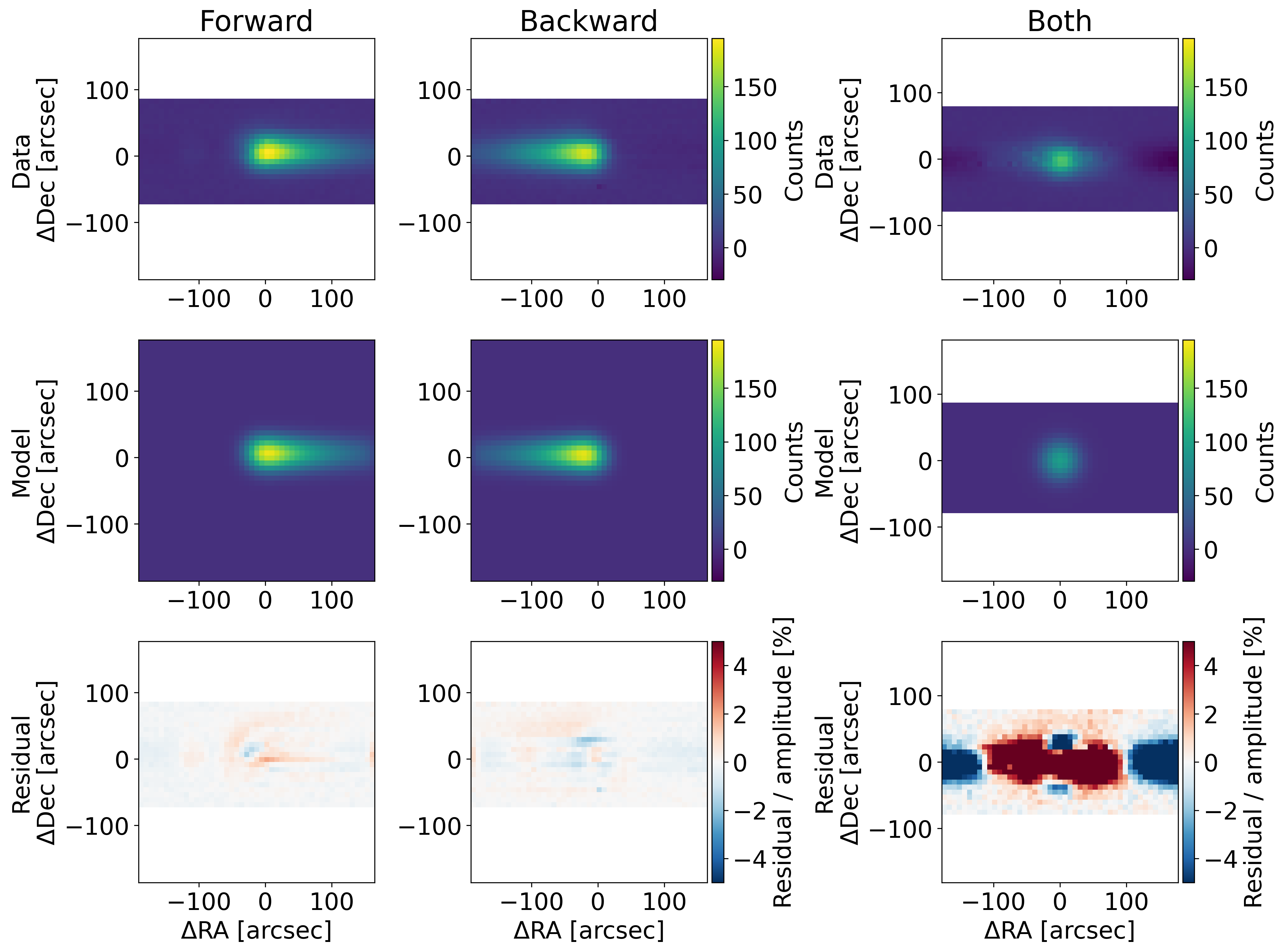}
    \caption{An example of the fitting process used for the joint forward and backward fits, as compared to the results obtained with a single Gaussian fit to the average of the two.}
    \label{fig:detector_fit_showcase}
\end{figure}

In the third step we split each detector image into a set of forward- and backward-scan maps producing two images per detector, as shown in Figure \ref{fig:detector_fit_showcase}. Since these types of detectors measure a temperature difference, there is a causal lag between the astrophysical source and the detector's response. The projected beam profile would then be measured as
\begin{equation}
    B = \mathcal{F}^{-1} \left( \mathcal{F}(G)\frac{1}{1 + i2\pi  k  s \alpha } \right).
    \label{eq:transfer_func}
\end{equation}
Above, $G$ represents a Gaussian kernel with center coordinates, $x_0$ and $y_0$, amplitude, $A$, and beam width, $\sigma$. We take $x$ to be along the scan direction and $y$ to be against, and the right hand term represents the effect of the causal lag where $s$ is $\pm1$ depending on scan direction, $k$ is the wavenumber, and $\alpha$ is the rate of decay and is converted to a time constant via the scan speed: $t_c = \alpha / v_s$. This beam profile is then convolved with a disk model for the observed planet, i.e. 
\begin{equation}
    B_m = \mathcal{F}^{-1}\left[\mathcal{F}(B)\mathcal{F}(D)\right]. 
\end{equation}
The disk is made by taking an empty grid and over sampling it at eight times higher resolution than the observation. We then emplace a flat disk of radius given by the planet's angular size as estimated from \citet{Park2021}. We then compute filling fraction of the over-sampled grid compared to the native resolution, which protects against aliasing and produces a more accurate measurement of the beam width.

The fitting procedure is then done as follows. We separate the data into forward- and backward-scan images for each detector. Then those data are jointly fit for the position of the source, its flux density, the detector's time constant, beam width, and an offset. Some detectors may have long tails due to slow time constants and so we use this information to first update the mask a third time.

In the fourth pass of this data we apply this new mask to the raw timestreams and then re-fit the joint forward- and backward-scan maps using the same joint fit procedure. From this group of fits we then extract a set of time constants that are converted to Hz using the telescope's scan speed ($\tau_c =  \frac{v_s}{\alpha}$) and are then used to inform the per-detector deconvolution.

In the final step we take the baseline subtracted timestreams from step four and divide out the transfer function of the time constant on a per-scan basis. This is schematically shown in Figure \ref{fig:scan_deconvolution}, where the transfer function for the time constant is a single pole roll off: 
\begin{equation}
    H(f) = \frac{1}{1 + i2\pi f/\tau_c}. 
    \label{eq:deconv_kernel}
\end{equation}
Above, $\tau_c$ is the time constant of the detector. We also employ the use of a low pass filter as this kernel systematically boosts high frequency data. The effect and choice of the low pass filter is discussed as applied to diffuse emission in \ref{sec:lpf}, where we terminate the filter with a sigmoid of width $2$~Hz centered at $10$~Hz. The end result of this deconvolution is that the amplitude of the planet is boosted and the long tail is removed as demonstrated in Figures \ref{fig:deconvolution_showcase} and \ref{fig:scan_deconvolution}. 

An extra effect becomes noticeable once deconvolving these data, which is that there is a delay between the forward and backward scans of around $-20$ to $-30$~ms which shifts the relative position of the planet in the forward- and backward-scan maps. We suspect that this is likely due to a constant group delay produced by the low-pass filter used in the readout electronics of the MCE box. We model this as
\begin{equation}
   D(f) = e^{-i2\pi f\Delta t}, 
\end{equation}
Since we expect this to be constant over an observation we add to the joint fits a fixed offset between the centers of the forward and backward maps equal to this group delay. For each planet observation the specific value of the group delay is tuned by repeating these fits over a set of values. We then use the pass 5 fits to determine the center of each detector in both the forward- and backward-scan maps. When the group delay is poorly tuned the fitted time constant absorbs part of it and as such we see a functional dependence on the difference of the along scan center. We find that these data are best described by a group delay of $-20$~ms or $-30$~ms which for a scan speed of $0.1$~deg/s translates to an offset of $14.4$ to $21.6$ arcseconds between the forward- and backward-scan maps.

The gains are estimated by comparing detectors to a fiducial model \citep{bean2022} similar to the scheme of \citet{Yang2026}, i.e. 
\begin{equation}
    S_{\nu} = B(\nu; T_b)\Omega.
\end{equation}
Above $\Omega$ is the solid angle subtended by the planet at the time of the observation which is estimated through the JPL ephemeris \citep{Park2021}. The gains are then computed as 
\begin{equation}
    G(\nu) = \frac{S_{\nu}}{A_c}.
    \label{eq:gain}
\end{equation}
Above, $A_c$ is the best fit amplitude in arbitrary data counts. We note that because the atmospheric opacity is corrected in an earlier step this methodology is consistent with the description provided by \citet{Yang2026}. We further compare the estimated gains from the deconvolved fits to a naive beam model + planet disk fit to the forward and backward scans of the second pass data and show that we are able to remove a tail of high gains, as seen in Figure \ref{fig:calibration}

\subsection{Applying Calibrations to OMC and G49.5}

Before any steps to process science data are made we apply the gain corrections and feedhorn offsets estimated from the planet observations. The feedhorn offset estimations are averaged over the set of planet observations, whereas the gain observations are attributed to science target data by matching observations that have similar atmospheric conditions and stage temperatures that were also taken at the same detector tuning parameters. The procedure to process these data then follows the flowchart outlined in Figure \ref{fig:flowchart}. We separate out these data into individual scans and remove baseline trends with a first order polynomial fit to the masked timestreams. The masks used are custom masks where we mask a circular region around the brightest point that encapsulates the visible diffuse emission of both sources. To match the way that the planet data are processed these data are deconvolved with the same per-detector kernel applied per-scan that is described in equation \ref{eq:deconv_kernel} and Figure \ref{fig:scan_deconvolution}. After deconvolution the per-scan variance is estimated as the population variance of that scan and added in quadrature with the gain uncertainty, i.e. 
\begin{equation}
    \sigma(t) = \left(V_s + \frac{D(t)^2}{G^2}\sigma_G^2\right)^{1/2}.
\end{equation}
Above, $V_s$ is the per-scan variance which is a constant over the whole scan, $D$ is the detector timestream over that scan, $G$ is the gain, and $\sigma_G$ is the uncertainty on the gain. These data are then projected onto a map using the same inverse variance weighted 2-D histogram procedure described in \S\ref{sec:map_making}. 

\subsection{The Effect of the Deconvolution Kernel on the Power Spectra}
\label{sec:lpf}

\begin{figure}
    \centering
    \includegraphics[width=1\linewidth]{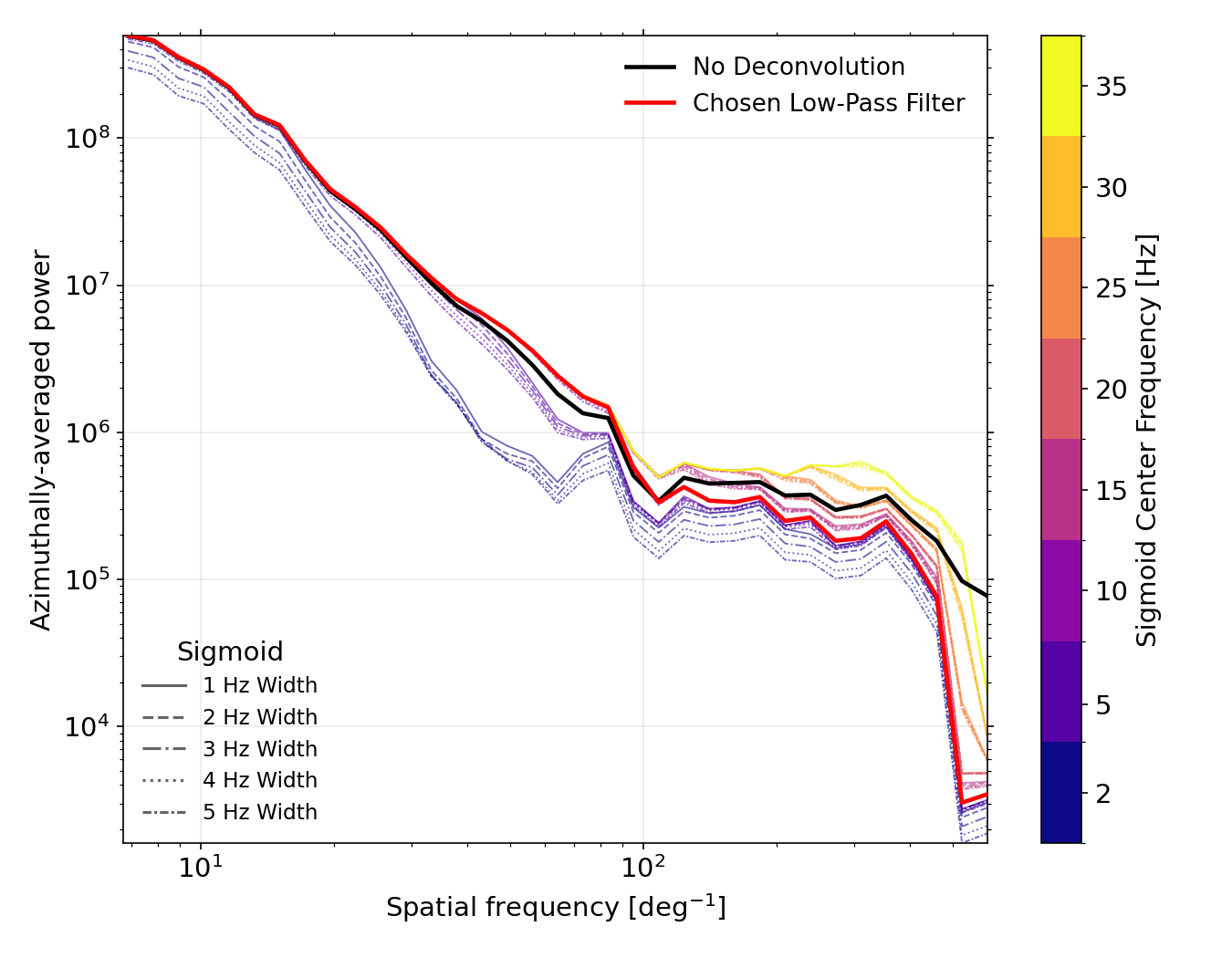}
    \caption{The radial power spectra of G49.5 as a function of the low-pass filter used in conjunction with the time constant deconvolution. The thick black line shows the baseline where there is no deconvolution or low pass filter applied. The thick redline shows the chosen low-pass filter that was used throughout this analysis. The idea behind this plot is to match the radial power spectra taken with these filters to the radial power spectrum without these filters i.e. we want to make sure we are removing as little astrophysical signal as possible. }
    \label{fig:deconvolution_noise}
\end{figure}

The deconvolution kernel is effective at recovering slow detectors that would otherwise have to be cut from the analysis. However, the kernel boosts high frequency noise, which if left alone can add unwanted artifacts into the images. To counteract this a low pass filter is included. We investigate here the choice of that filter by taking a slow detector ($8$~Hz) from the G49.5 data set and computing the radial power spectrum at a set of low pass filter configurations. Each filter has a value of one over allowable frequencies and is terminated using a sigmoid function centered with a tunable frequency that ramps down over a tunable width. From the analysis shown in Figure \ref{fig:deconvolution_noise} we opt to use a filter with a center of $10$~Hz and a width of $2$~Hz as it most closely recovers the power spectral density when no deconvolution is applied. 

\begin{figure}
    \centering
    \includegraphics[width=1\linewidth]{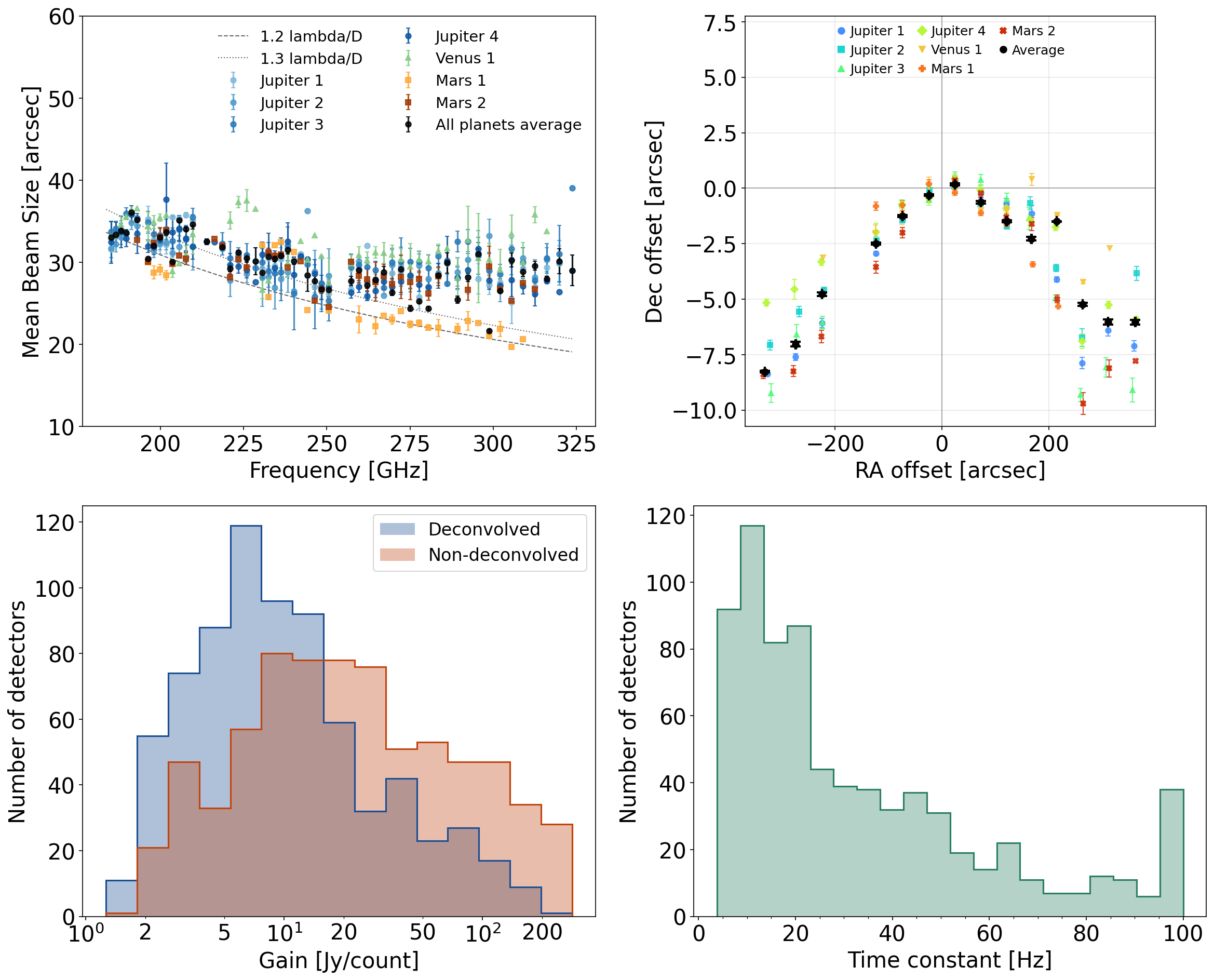}
    \caption{A summary plot of the relevant fitted values over the set of planet observations used as calibrators in this analysis. The top left panel gives the FWHM values of the beam, the top right gives the relative feedhorn offsets, the bottom left panel shows a histogram of the gains, and the bottom right shows a histogram of the time constant. A key take-away from the gains plot is that we are able to correct for the high-gain tail via the deconvolution with the time constants. }
    \label{fig:calibration}
\end{figure}

\subsection{Quality Checks on OMC and G49.5 Images}


Now that we have a set of calibrated maps of both OMC and G49.5 we go through another set of quality checks to keep high quality detectors. The first of these checks is to remove the edge channels. TIME was intentionally designed to have a pass band between two water vapor lines, where the lowest and highest frequency channels are explicitly observing the atmosphere.

We manually cut detectors that have significant noise visible in their individual images as well as detectors that have bright systematics present in the data. We then take the inverse variance weighting of these data per spectral channel. 

\section{G49.5 and OMC Observations}
\label{sec:g49}

The BN-KL region of the Orion Molecular Cloud was observed at least once a day during TIME commissioning operations for a total of $20$ observations, and over a range of scan parameters, shown in Table \ref{tab:obs_table}. In addition to this, the detectors were tuned three times throughout these operations giving us three different calibrations. This makes OMC a good testbed for understanding the stability of our instrument. 

G49.5 is a bright HII region in the Milky Way \citep{Carpenter_1998} and is a massive star forming complex that consists of multiple smaller clouds \citep{lim_2019}. We observed one of the smaller clouds, G49.5, during the TIME 2022 commissioning run, and processed it according to the procedure outlined in \S\ref{sec:map_making}. 
\subsection{G49.5}

We compare TIME observations to Bolocam observations \citep{bgps} of G49.5 using the following methodology. The individual spectral channels of TIME are convolved to a common beam size, which is estimated as the largest FWHM ($37$~arcsec) in the fitted values in the top left panel of Figure \ref{fig:calibration}. Bolocam, which has a beam size of $33$~arcsec, is also convolved to the common beam size of $37$~arcseconds. 
\begin{equation}
    F_{BB}(x,y) = \frac{\sum_{\nu} F(\nu,x,y)T(\nu)}{ \sum_{\nu}T(\nu)}.
    \label{eq:broadband_add}
\end{equation}
Above, $F_{BB}$ is the flux density of the broadband image, $\nu$ is the frequency dependence, $x$ and $y$ are the pixel coordinates of the image, and $T(\nu)$ is the spectral response function of Bolocam \citep{bgps,Ginsburg2013}. TIME is missing spectral channels within Bolocam's passband that would account for a total of $22\%$ of the flux density observed. When computing broadband TIME image we include only the bands observed by TIME and remove this percentage of flux from the Bolocam image. We isolate pixel covariance by sampling the image at points separated by one width in both $x$ and $y$ coordinates and fit a straight line with a fixed offset of zero. We find that we are within $2\%$ calibration of Bolocam, which had a model uncertainty of $5\%$ for the calibrators used by them.
\begin{figure}[h!]
    \centering
    \includegraphics[width=1\linewidth]{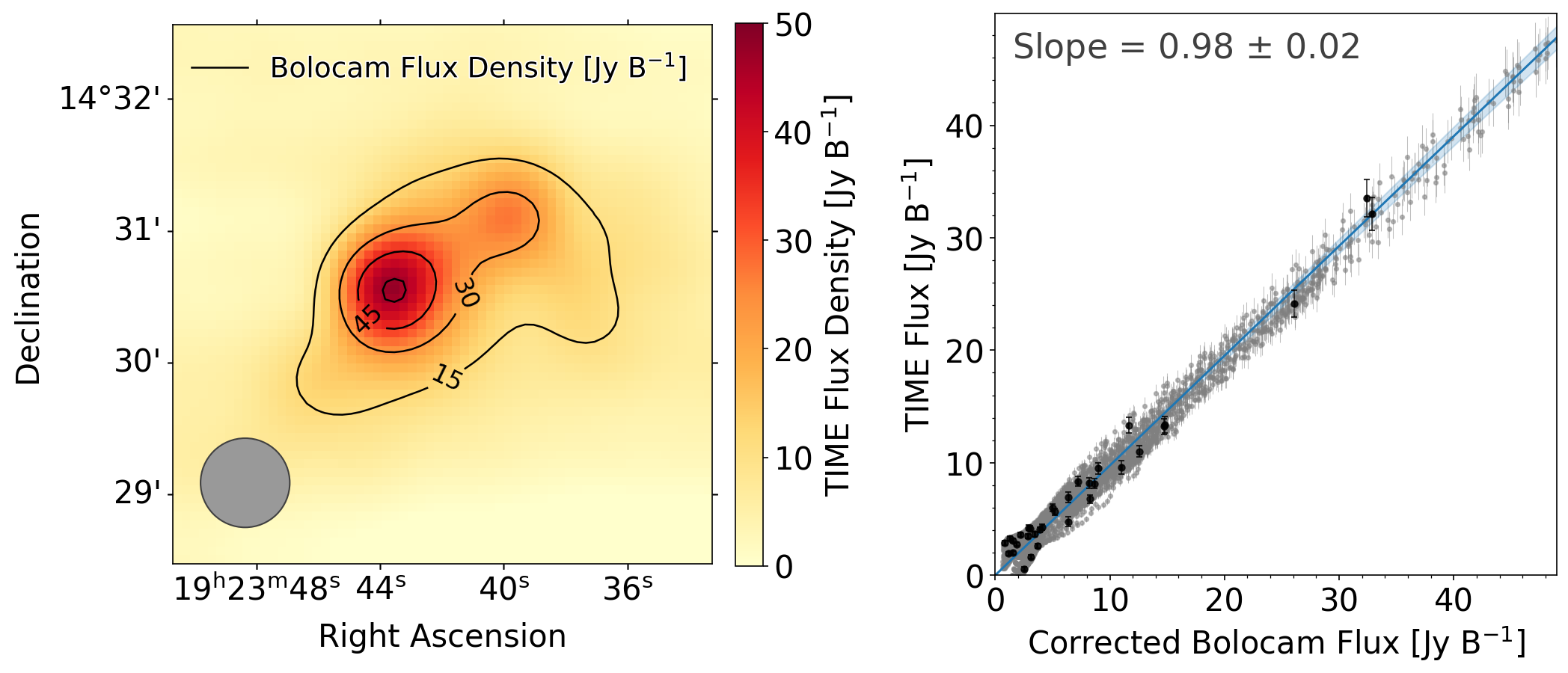}
    \caption{The left panel shows a broadband TIME image of G49.5, estimated by co-adding spectral channels of TIME using equation \ref{eq:broadband_add}, there are Bolocam contours overlaid which demonstrate that the pointing is the same. On the right is the a pixel-to-pixel plot of the TIME image compared to Bolocam. The semi-transparent gray points show the distribution of every pixel, and the black solid points indicate a set of $36$ points that are separated by the beam width of the instrument. As such these black points allow us to remove pixel covariance where all the pixels inside a beam size are correlated. A line is fit to these black points and we show that we are within $2\%$ flux calibration of Bolocam, which is less than the $5\%$ model uncertainty of the ALMA calibrators.}
    \label{fig:bgps_comp}
\end{figure}

\subsection{OMC}
\label{sec:OMC}

We have also processed observations of the OMC complex which was observed at least once per day through the 2022 winter commissioning. Of the $20$ OMC observations taken during this run we limit our analysis to just the six that were taken in the some detector tuning period as the G49.5 observations. Those data are processed using the procedure described in \S\ref{sec:map_making} in the same manner as the G49.5 observations.

\begin{figure}[h!]
    \centering
    \includegraphics[width=1\linewidth]{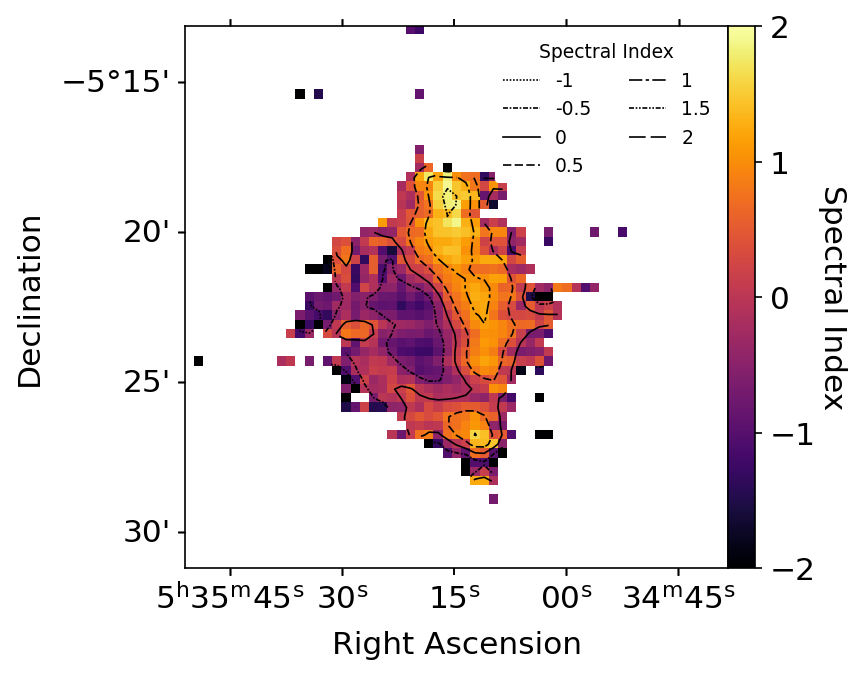}
    \caption{A spectral index map of OMC, with contour lines of spectral indices overlaid. The spectral index of the BN/KL region is between $1$ and $1.5$ which is consistent with the expectation for a star forming region, and other measurements \citep{sert2025}.}
    \label{fig:OMC_spectral_indexes}
\end{figure}

\makeatletter\@booleanfalse\force@deferlist@sw\makeatother

\begin{figure*}
    \centering
    \includegraphics[width=1\linewidth]{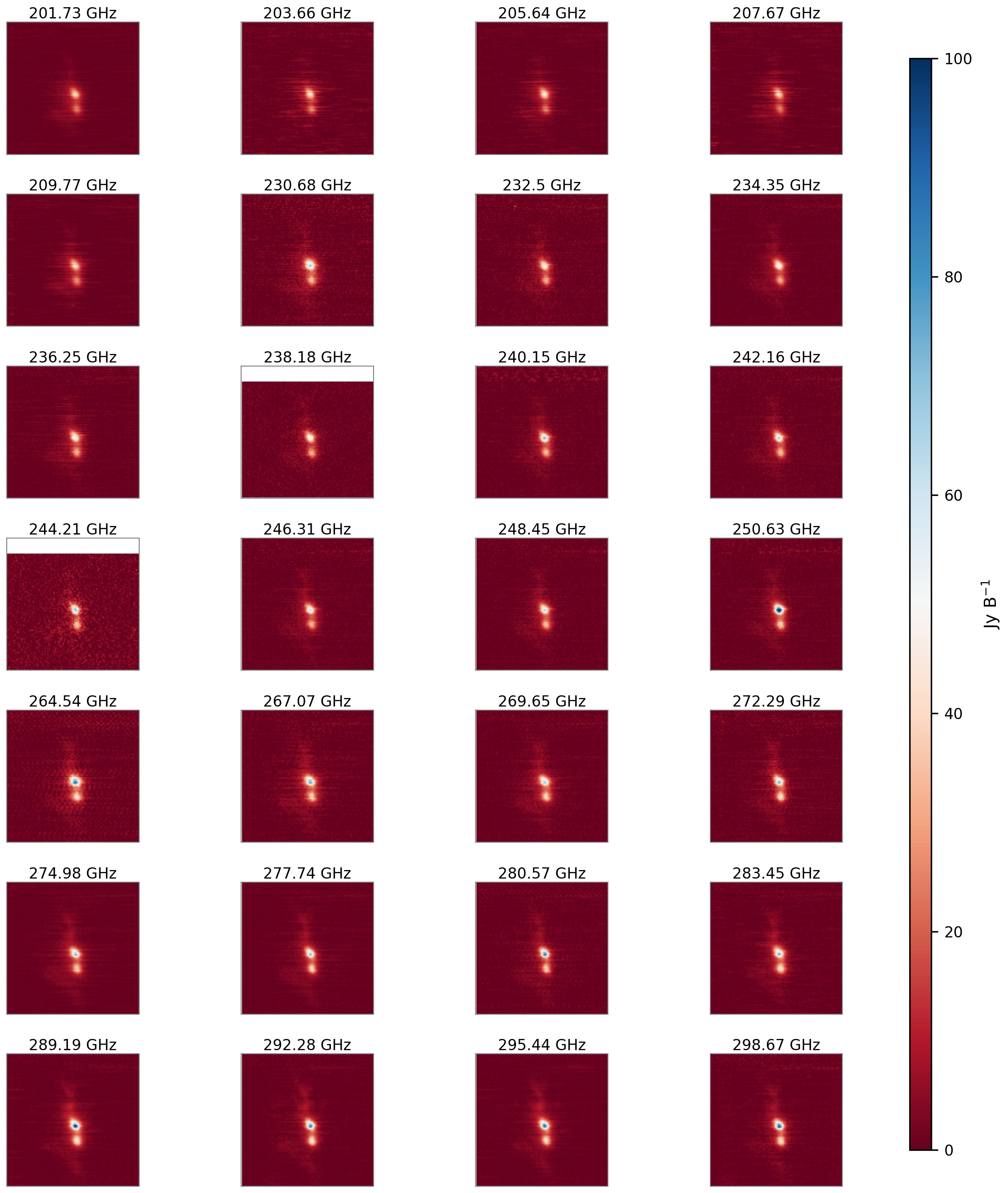}
    \caption{The spectral images of OMC after we have averaged across a sub-set of OMC observations that were taken at a smaller raster step size, as denoted in Table \ref{tab:obs_table}, which allows for them to make better use of the spatial resolution available at this facility. }
    \label{fig:OMC_data}
\end{figure*}

\begin{deluxetable*}{lccc}[t!]
    \tablecaption{Sensitivity and Mapping Speed Estimates}
    \label{tab:sensitivity_table}
    \tablehead{\colhead{target signal} &\colhead{LIM field} & \colhead{SZe} & \colhead{Pre-Stellar Core}}
    \startdata 
    Required Field Size & $1.3\arcdeg \times 0.007\arcdeg$ & $0.23\arcdeg \times 0.08\arcdeg$ & $0.23\arcdeg \times 0.08\arcdeg$ \\ 
    Desired Sensitivity & $0.02$~MJy~sr$^{-1}$ & $0.04$~MJy~sr$^{-1}$ & $50$~mJy \\ 
    Hours Required & $403$ & $208$ & $3$ 
    \enddata
    \tablecomments{TIME is designed to observe a 1-D patch of sky. As such, we demonstrate the total number of observing time needed to integrate down a $1.3\arcdeg$ line (as simulated in \citet{Sun2021}) to reach a sensitivity of $0.02$~MJy sr$^{-1}$ \citep{Chung2020} is $403$~hours.}
\end{deluxetable*}

To produce a set of spectra for the OMC BN-KL region we use the technique of aperture photometry.

In this scheme we take the sum of the flux density of all pixels within $x$~arcseconds of the center of BN-KL. This is done over the set of observations in Group C of Table \ref{tab:obs_table}. We use only this group of observations as it contains the OMC observations that were taken within the same detector tuning as the G49.5 observations where we compare the absolute flux calibration.

We also generate a map of the spectral indices. To do this we average these observations together by frequency and fit a power law to each pixel of the image. Frequencies in the range of $254-265$~GHz are masked so that we do not bias these fits with spectral line emission.  The resulting spectral index map is shown in Figure \ref{fig:OMC_spectral_indexes}.

\begin{figure}
    \centering
    \includegraphics[width=1\linewidth]{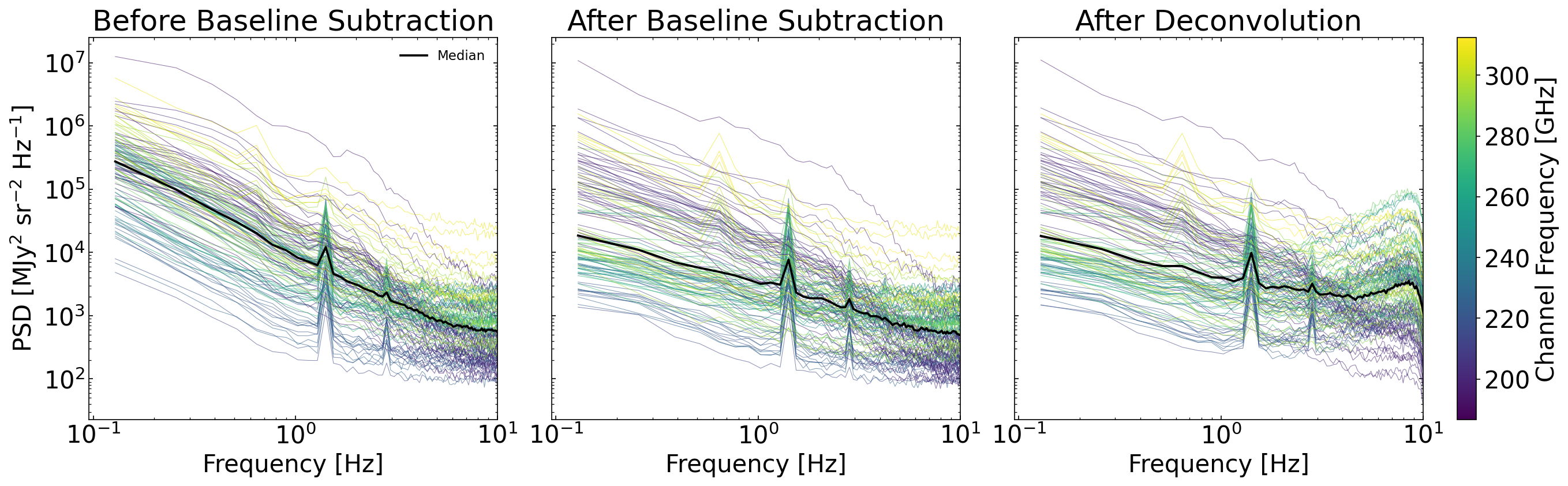}
    \caption{Power spectra of a single OMC observation. The different lines indicate different detectors and they are colored by the frequency of their spectral channel. These power spectra are computed over off-source scans that do not contain bright emission from OMC. The left shows the PSD of the raw timestreams, the middle panel shows the PSD after a first order polynomial has been fit and subtracted, and the right shows the PSD after the detectors have been deconvolved by the time constant kernel which causes an increase at high frequencies. }
    \label{fig:PSD}
\end{figure}

A set of off-source scans are used to gauge the sensitivity of these data. To do this we compute the power spectral density (PSD) of each off source scan for each detector. Then per-detector we compute the median PSD. This is shown in Figure \ref{fig:PSD} where we show the PSD at a few different stages of the analysis. The sensitivity is then estimated as the median of the PSD over the range of $3$ to $10$~Hz. We average these per detector over the set of six observations, and then further average detectors within the same spectral channel.

\section{Discussion}

\label{sec:discussion}

We now discuss some lessons learned and improvements that could be made to the TIME instrument for upcoming observing runs. Inhomogeneities inherent in the small-scale detector fabrication associated with a pathfinder instrument on the scale of TIME, will require particular care in future observing operations. All else being equal, achieved on-sky detector gains will change with both atmospheric loading and the thermal bath temperature of the focal plane, both disturbing the overall thermoelectric feedback loop that governs the equilibrium state of the TES and altering the responsivity of the detector.

In future observing runs we will use proportional/integral/derivative (PID) loop to stabilize the stage temperature and remove this contribution to bias point drift that may affect gains and yield. This system was prototyped during the 2022 commissioning observations, but due to temperature instabilities from aging equipment it reduced the amount of observing time available each day and we thus decided to turn it off. These old parts are currently being refurbished and we anticipate in future observing seasons we will be able to consistently PID the focal plane and reduce the effect of thermal drifts on gain sensitivity, which is crucial to probing the faint LIM signals we are interested in observing. 

\begin{figure}[h!]
    \centering
    \includegraphics[width=1\linewidth]{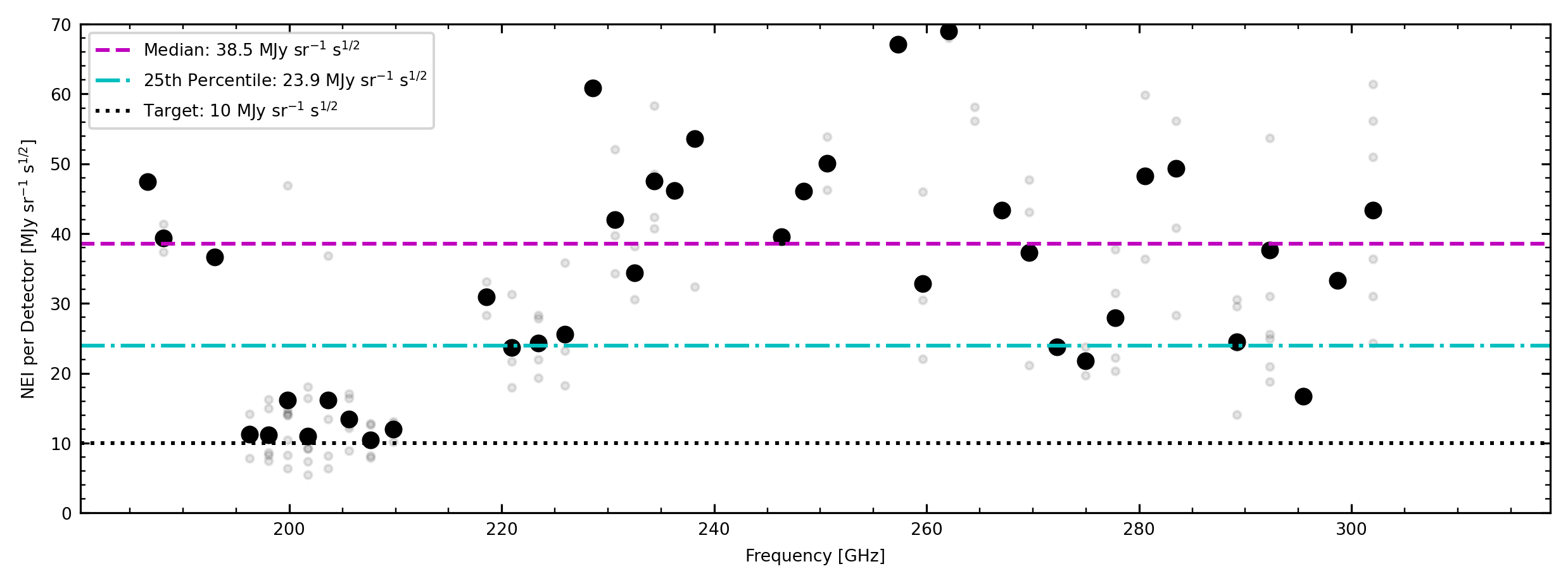}
    \caption{The per-detector NEI estimated by power spectra of off source scans, such as the ones shown in Figure \ref{fig:PSD}. The sensitivity here is estimated at the range of $3$ to $10$~Hz. The gray points show each detector, and the black points show inverse NEI$^2$ weighted average of each detector within a given spectral channel.  }
    \label{fig:sensitivity_plots}
\end{figure}

\begin{figure*}
    \centering
    \includegraphics[width=1\linewidth]{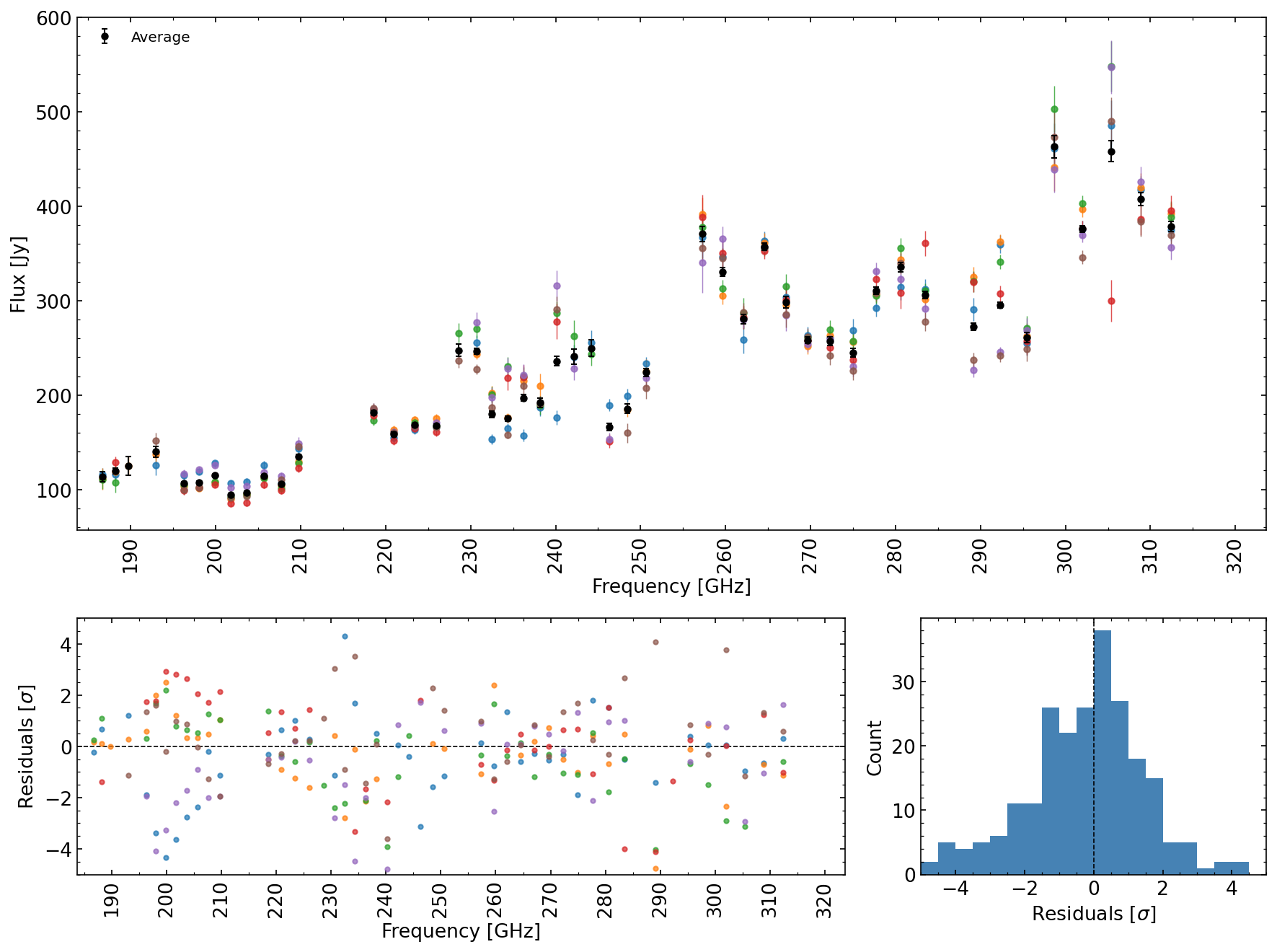}
    \caption{Spectra of the BN/KL region of OMC. Different observations are recorded in different colors with the inverse variance weighted average of these observations in black. In the bottom panels a residual plot is shown where we estimate the difference between each observation and the averaged spectrum in units of uncertainty. To the right is a histogram of the residuals, also in units of uncertainty. Some channels show differences in excess of what we would expect from statistical uncertainty, sources of these systematics are discussed in section \ref{sec:discussion}. Missing channels are omitted where there were no detectors that were sensitive during the observation.}
    \label{fig:OMC_spectra}
\end{figure*}
\makeatletter\@booleantrue\force@deferlist@sw\makeatother

We are also working towards building detectors that can be more stably biased at a common set point \citep{TIME_TES}. Beyond that, we now have many more detector modules than in 2022. The modular nature of the focal plane allows us to better place these modules in a way that maximizes on-sky yield.

With the opacity at Kitt Peak being fairly high ($\tau\approx0.1$ at zenith in top 10\% conditions during the 2022 observing season), real-time monitoring of the atmosphere is more important than at other high and dry sites. During the 2022 run the dedicated tipping radiometer at the ARO facility was non-operational, forcing us to rely purely on atmospheric forecasting models. However, for our subsequent 2026 engineering run this radiometer was operational alongside a newly developed weather station. We are investigating the possible long-term use of this weather station alongside global navigation satellite system signals for high-fidelity, cost-effective estimates of precipitable water vapor content (Zemcov et al., in prep; see also previous work by~\cite{2022AJ....163..283W,2024MNRAS.528.4582S}). In addition to these improvements of opacity monitoring on the site, we will be making empirical measurements of the atmosphere during our scientific observing runs.

\subsection{Expectations for Future TIME Sensitivities}

We estimate the aggregate sensitivity of the TIME instrument to inform future scientific deployments. This quantity is derived directly from the per-scan variance of each detector in each observation. Subsequent scans in the same observation are averaged together, we then further average observations together on a per-detector basis. Due to the prototype nature of the 2022 observing run, we use the top performing detectors as a measure of the achievable sensitivities with a more developed instrument. These values are $300$~mJy\,s$^{1/2}$ or $10$~MJy\,sr$^{-1}$s$^{1/2}$ per detector. Simulations from \citet{Sun2021} estimate the constraining power we can achieve on different [CII] and CO power spectra observables assuming a noise equivalent intensity of $5$~MJy\,sr$^{-1}$\,s$^{1/2}$. Thus demonstrating that a more mature instrument with accurate improved noise removal will be capable of measuring LIM signals. This sensitivity would also allow us to access small surveys to perform additional science cases such as the SZ effect in galaxy clusters, pre-stellar cores, and planetary nebulae in hundreds of hours of integration time, as tabulated in Table \ref{tab:}. 

We have built a raw data processing pipeline that produces calibrated spectral images of any science target with TIME. This pipeline was then used to produce spectral images of OMC and G49.5. Then, we assessed the relative calibration of TIME to another instrument and to itself with subsequent observations. Based on these outcomes we have demonstrated a preliminary performance of the TIME instrument and outlined a path forward for improving instrument performance.

\section{Acknowledgments}
This material is based upon work supported by the National Science Foundation under Grant Nos.~1602677, 1653228, 1910598, 2308039, 2308040, 2308041, and 2308042.

The TIME collaboration acknowledges the critical contributions of the Arizona Radio Observatory engineers and operators, including Kevin Bays, Tom Folkers, Natalie Gandilo, Blythe Guvenen, Sean Keel, and Martin McColl. We also acknowledge Sukhman Singh and Ibrahim Shehzad at Cornell and Lisa Nasu-Yu at the University of Toronto for early contributions to data analysis, observations, and simulations that ultimately informed this work.

ATC acknowledges support as a Fred Young Faculty Fellow at Cornell University.

DTC was supported by a CITA/Dunlap Institute postdoctoral fellowship for much of this work. The Dunlap Institute is funded through an endowment established by the David Dunlap family and the University of Toronto. Research in Canada is supported by NSERC and CIFAR.

We acknowledge the use of Claude Code for its help in producing figures and some high level analysis software. All code that is made through generative AI is still read, debugged, tested, and run by BJV.

\bibliography{sample701}{}

@INPROCEEDINGS{TIME_pilot,
       author = {{Crites}, A.~T. and {Bock}, J.~J. and {Bradford}, C.~M. and {Chang}, T.~C. and {Cooray}, A.~R. and {Duband}, L. and {Gong}, Y. and {Hailey-Dunsheath}, S. and {Hunacek}, J. and {Koch}, P.~M. and {Li}, C.~T. and {O'Brient}, R.~C. and {Prouve}, T. and {Shirokoff}, E. and {Silva}, M.~B. and {Staniszewski}, Z. and {Uzgil}, B. and {Zemcov}, M.},
        title = "{The TIME-Pilot intensity mapping experiment}",
    booktitle = {Millimeter, Submillimeter, and Far-Infrared Detectors and Instrumentation for Astronomy VII},
         year = 2014,
       editor = {{Holland}, Wayne S. and {Zmuidzinas}, Jonas},
       series = {Society of Photo-Optical Instrumentation Engineers (SPIE) Conference Series},
       volume = {9153},
        month = aug,
          eid = {91531W},
        pages = {91531W},
          doi = {10.1117/12.2057207},
       adsurl = {https://ui.adsabs.harvard.edu/abs/2014SPIE.9153E..1WC}
}

@ARTICLE{TIME_TES,
  author={Butler, Victoria L. and Bock, James J. and Chung, Dongwoo T. and Crites, Abigail T. and Frez, Clifford and Greathouse, Annsley and Lau, King and Lowe, Ian and Marrone, Dan P. and Mayer, Evan C. and Vaughan, Benjamin J. and Zemcov, Michael},
  journal={IEEE Transactions on Applied Superconductivity}, 
  title={TES Bolometer Design and Testing for the Tomographic Ionized-Carbon Mapping Experiment Millimeter Array}, 
  year={2026},
  volume={36},
  number={6},
  pages={2103507-2103507},
  doi={10.1109/TASC.2026.3694584}}

@ARTICLE{Battiselli_2008,
       author = {{Battistelli}, E.~S. and {Amiri}, M. and {Burger}, B. and {Halpern}, M. and {Knotek}, S. and {Ellis}, M. and {Gao}, X. and {Kelly}, D. and {Macintosh}, M. and {Irwin}, K. and {Reintsema}, C.},
        title = "{Functional Description of Read-out Electronics for Time-Domain Multiplexed Bolometers for Millimeter and Sub-millimeter Astronomy}",
      journal = {Journal of Low Temperature Physics},
         year = 2008,
        month = may,
       volume = {151},
       number = {3-4},
        pages = {908-914},
          doi = {10.1007/s10909-008-9772-z},
       adsurl = {https://ui.adsabs.harvard.edu/abs/2008JLTP..151..908B}
}

@ARTICLE{bgps,
       author = {{Aguirre}, James E. and {Ginsburg}, Adam G. and {Dunham}, Miranda K. and {Drosback}, Meredith M. and {Bally}, John and {Battersby}, Cara and {Bradley}, Eric Todd and {Cyganowski}, Claudia and {Dowell}, Darren and {Evans}, II, Neal J. and {Glenn}, Jason and {Harvey}, Paul and {Rosolowsky}, Erik and {Stringfellow}, Guy S. and {Walawender}, Josh and {Williams}, Jonathan P.},
        title = "{The Bolocam Galactic Plane Survey: Survey Description and Data Reduction}",
      journal = {\apjs},
         year = 2011,
        month = jan,
       volume = {192},
       number = {1},
          eid = {4},
        pages = {4},
          doi = {10.1088/0067-0049/192/1/4},
archivePrefix = {arXiv},
       eprint = {1011.0691},
 primaryClass = {astro-ph.GA},
       adsurl = {https://ui.adsabs.harvard.edu/abs/2011ApJS..192....4A}
}

@inproceedings{SPIE_proceeding,
author = {Victoria L. Butler and Abigail T. Crites and Samantha Berek and James J. Bock and Geoff Bower and C. Matt Bradford and Tessalie-Caze Cortez and Tzu-Ching Chang and Yun-Ting Cheng and Dongwoo T. Chung and Asantha Corray and Audrey Dunn and Nick Emerson and Clifford Frez and Minal Shaik and Francie Wharton and Caidan Pilarski and Sarah Gates and Caleb Greenburg and Fiona Hufford and Jonathon Hunacek and Ryan P. Keenan and Baria Khan and King Lau and Chao-Te Li and Ian Lowe and Paolo Madonia and Dan P. Marrone and Evan C. Mayer and Lorenzo Moncelsi and Sofia Pereira and Dang Pham and Ibrahim Shehzad and Sukhman Singh and Guochao Sun and Isaac Trumper and Anthony Turner and Benjamin J. Vaughan and Tashun Wei and Quinn Wilson and Michael Zemcov},
title = {{TIME: the Tomographic Ionized-carbon Mapping Experiment: an update on design, characterization, and data from the 2022 commissioning observations}},
volume = {13102},
booktitle = {Millimeter, Submillimeter, and Far-Infrared Detectors and Instrumentation for Astronomy XII},
editor = {Jonas Zmuidzinas and Jian-Rong Gao},
organization = {International Society for Optics and Photonics},
publisher = {SPIE},
pages = {131022G},
year = {2024},
doi = {10.1117/12.3021442},
URL = {https://doi.org/10.1117/12.3021442}
}

@ARTICLE{Mayer_2026,
       author = {{Mayer}, Evan C. and {Lowe}, Ian N. and {Marrone}, Daniel P. and {Bock}, James J. and {Bradford}, Charles M. and {Butler}, Victoria L. and {Chang}, Tzu-Ching and {Cheng}, Yun-Ting and {Chung}, Dongwoo T. and {Crites}, Abigail T. and {Dunn}, Audrey and {Emerson}, Nicholas and {Frez}, Clifford and {Hunacek}, Jonathon and {Keenan}, Ryan P. and {Li}, Chao-Te and {Lau}, King and {Sun}, Guochao and {Trumper}, Isaac and {Turner}, Anthony D. and {Vaughan}, Benjamin and {Wei}, Ta-Shun and {Zemcov}, Michael},
        title = "{Development of a planar cable-driven parallel robot for submillimeter and terahertz beam mapping measurements}",
      journal = {arXiv e-prints},
         year = 2025,
        month = nov,
          eid = {arXiv:2511.09446},
        pages = {arXiv:2511.09446},
          doi = {10.48550/arXiv.2511.09446},
archivePrefix = {arXiv},
       eprint = {2511.09446},
 primaryClass = {astro-ph.IM},
       adsurl = {https://ui.adsabs.harvard.edu/abs/2025arXiv251109446M}
}

@ARTICLE{sert2025,
       author = {{D{\'e}sert}, F.-X. and {Mac{\'\i}as-P{\'e}rez}, J.~F. and {Beelen}, A. and {Beno{\^\i}t}, A. and {B{\'e}thermin}, M. and {Bounmy}, J. and {Bourrion}, O. and {Calvo}, M. and {Catalano}, A. and {De Breuck}, C. and {Dubois}, C. and {Dur{\'a}n}, C.~A. and {Fasano}, A. and {Goupy}, J. and {Hu}, W. and {Ibar}, E. and {Lagache}, G. and {Lundgren}, A. and {Monfardini}, A. and {Ponthieu}, N. and {Quinatoa}, D. and {Van Cuyck}, M. and {Adam}, R. and {Ade}, P. and {Ajeddig}, H. and {Amarantidis}, S. and {Andr{\'e}}, P. and {Aussel}, H. and {Berta}, S. and {Bongiovanni}, A. and {Ch{\'e}rouvrier}, D. and {De Petris}, M. and {Doyle}, S. and {Driessen}, E.~F.~C. and {Ejlali}, G. and {Ferragamo}, A. and {Gomez}, A. and {Hanser}, C. and {Katsioli}, S. and {K{\'e}ruzor{\'e}}, F. and {Kramer}, C. and {Ladjelate}, B. and {Leclercq}, S. and {Lestrade}, J.-F. and {Madden}, S.~C. and {Maury}, A. and {Mayet}, F. and {Moyer-Anin}, A. and {Mu{\~n}oz-Echeverr{\'\i}a}, M. and {Myserlis}, I. and {Paliwal}, A. and {Perotto}, L. and {Pisano}, G. and {Rev{\'e}ret}, V. and {Rigby}, A.~J. and {Ritacco}, A. and {Roussel}, H. and {Ruppin}, F. and {S{\'a}nchez-Portal}, M. and {Savorgnano}, S. and {Sievers}, A. and {Tucker}, C. and {Zylka}, R.},
        title = "{Continuum, CO, and water vapour maps of the Orion Nebula: First millimetre spectral imaging with CONCERTO}",
      journal = {\aap},
         year = 2025,
        month = sep,
       volume = {701},
          eid = {A210},
        pages = {A210},
          doi = {10.1051/0004-6361/202555320},
archivePrefix = {arXiv},
       eprint = {2504.20487},
 primaryClass = {astro-ph.GA},
       adsurl = {https://ui.adsabs.harvard.edu/abs/2025A&A...701A.210D}
}

@ARTICLE{Sun2021,
       author = {{Sun}, G. and {Chang}, T.-C. and {Uzgil}, B.~D. and {Bock}, J.~J. and {Bradford}, C.~M. and {Butler}, V. and {Caze-Cortes}, T. and {Cheng}, Y.-T. and {Cooray}, A. and {Crites}, A.~T. and {Hailey-Dunsheath}, S. and {Emerson}, N. and {Frez}, C. and {Hoscheit}, B.~L. and {Hunacek}, J. and {Keenan}, R.~P. and {Li}, C.~T. and {Madonia}, P. and {Marrone}, D.~P. and {Moncelsi}, L. and {Shiu}, C. and {Trumper}, I. and {Turner}, A. and {Weber}, A. and {Wei}, T.~S. and {Zemcov}, M.},
        title = "{Probing Cosmic Reionization and Molecular Gas Growth with TIME}",
      journal = {\apj},
         year = 2021,
        month = jul,
       volume = {915},
       number = {1},
          eid = {33},
        pages = {33},
          doi = {10.3847/1538-4357/abfe62},
archivePrefix = {arXiv},
       eprint = {2012.09160},
 primaryClass = {astro-ph.GA},
       adsurl = {https://ui.adsabs.harvard.edu/abs/2021ApJ...915...33S}
}

@software{Paine2019,
       author = {{Paine}, Scott},
        title = "{am: Microwave through submillimeter-wave propagation tool for the terrestrial atmosphere}",
 howpublished = {Astrophysics Source Code Library, record ascl:2205.002},
         year = 2022,
        month = may,
          eid = {ascl:2205.002},
archivePrefix = {ascl},
       eprint = {2205.002},
       adsurl = {https://ui.adsabs.harvard.edu/abs/2022ascl.soft05002P}
}

@article{bean2022,
       author = {{CASA Team} and {Bean}, Ben and {Bhatnagar}, Sanjay and {Castro}, Sandra and {Donovan Meyer}, Jennifer and {Emonts}, Bjorn and {Garcia}, Enrique and {Garwood}, Robert and {Golap}, Kumar and {Gonzalez Villalba}, Justo and {Harris}, Pamela and {Hayashi}, Yohei and {Hoskins}, Josh and {Hsieh}, Mingyu and {Jagannathan}, Preshanth and {Kawasaki}, Wataru and {Keimpema}, Aard and {Kettenis}, Mark and {Lopez}, Jorge and {Marvil}, Joshua and {Masters}, Joseph and {McNichols}, Andrew and {Mehringer}, David and {Miel}, Renaud and {Moellenbrock}, George and {Montesino}, Federico and {Nakazato}, Takeshi and {Ott}, Juergen and {Petry}, Dirk and {Pokorny}, Martin and {Raba}, Ryan and {Rau}, Urvashi and {Schiebel}, Darrell and {Schweighart}, Neal and {Sekhar}, Srikrishna and {Shimada}, Kazuhiko and {Small}, Des and {Steeb}, Jan-Willem and {Sugimoto}, Kanako and {Suoranta}, Ville and {Tsutsumi}, Takahiro and {van Bemmel}, Ilse M. and {Verkouter}, Marjolein and {Wells}, Akeem and {Xiong}, Wei and {Szomoru}, Arpad and {Griffith}, Morgan and {Glendenning}, Brian and {Kern}, Jeff},
        title = "{CASA, the Common Astronomy Software Applications for Radio Astronomy}",
      journal = {\pasp},
         year = 2022,
        month = nov,
       volume = {134},
       number = {1041},
          eid = {114501},
        pages = {114501},
          doi = {10.1088/1538-3873/ac9642},
archivePrefix = {arXiv},
       eprint = {2210.02276},
 primaryClass = {astro-ph.IM},
       adsurl = {https://ui.adsabs.harvard.edu/abs/2022PASP..134k4501C}
}

@ARTICLE{Park2021,
       author = {{Park}, Ryan S. and {Folkner}, William M. and {Williams}, James G. and {Boggs}, Dale H.},
        title = "{The JPL Planetary and Lunar Ephemerides DE440 and DE441}",
      journal = {\aj},
         year = 2021,
        month = mar,
       volume = {161},
       number = {3},
          eid = {105},
        pages = {105},
          doi = {10.3847/1538-3881/abd414},
       adsurl = {https://ui.adsabs.harvard.edu/abs/2021AJ....161..105P}
}

@ARTICLE{2024MNRAS.528.4582S,
       author = {{Sugiyama}, Junna and {Nishino}, Haruki and {Kusaka}, Akito},
        title = "{Precipitable water vapour measurement using GNSS data in the Atacama Desert for millimetre and submillimetre astronomical observations}",
      journal = {\mnras},
         year = 2024,
        month = mar,
       volume = {528},
       number = {3},
        pages = {4582-4590},
          doi = {10.1093/mnras/stae270},
archivePrefix = {arXiv},
       eprint = {2308.12632},
 primaryClass = {astro-ph.IM},
       adsurl = {https://ui.adsabs.harvard.edu/abs/2024MNRAS.528.4582S}
}

@ARTICLE{2022AJ....163..283W,
       author = {{Wood-Vasey}, W.~M. and {Perrefort}, Daniel and {Baker}, Ashley D.},
        title = "{GPS Measurements of Precipitable Water Vapor Can Improve Survey Calibration: A Demonstration from KPNO and the Mayall z-band Legacy Survey}",
      journal = {\aj},
         year = 2022,
        month = jun,
       volume = {163},
       number = {6},
          eid = {283},
        pages = {283},
          doi = {10.3847/1538-3881/ac63bb},
       adsurl = {https://ui.adsabs.harvard.edu/abs/2022AJ....163..283W}
}

@ARTICLE{Chung2020,
       author = {{Chung}, Dongwoo T. and {Viero}, Marco P. and {Church}, Sarah E. and {Wechsler}, Risa H.},
        title = "{Forecasting [C II] Line-intensity Mapping Measurements between the End of Reionization and the Epoch of Galaxy Assembly}",
      journal = {\apj},
         year = 2020,
        month = mar,
       volume = {892},
       number = {1},
          eid = {51},
        pages = {51},
          doi = {10.3847/1538-4357/ab798f},
archivePrefix = {arXiv},
       eprint = {1812.08135},
 primaryClass = {astro-ph.GA},
       adsurl = {https://ui.adsabs.harvard.edu/abs/2020ApJ...892...51C}
}

@ARTICLE{kovetz_2019,
       author = {{Kovetz}, Ely and {Breysse}, Patrick C. and {Lidz}, Adam and {Bock}, Jamie and {Bradford}, Charles M. and {Chang}, Tzu-Ching and {Foreman}, Simon and {Padmanabhan}, Hamsa and {Pullen}, Anthony and {Riechers}, Dominik and {Silva}, Marta B. and {Switzer}, Eric},
        title = "{Astrophysics and Cosmology with Line-Intensity Mapping}",
      journal = {\baas},
         year = 2019,
        month = may,
       volume = {51},
       number = {3},
          eid = {101},
        pages = {101},
          doi = {10.48550/arXiv.1903.04496},
archivePrefix = {arXiv},
       eprint = {1903.04496},
 primaryClass = {astro-ph.CO},
       adsurl = {https://ui.adsabs.harvard.edu/abs/2019BAAS...51c.101K}
}

@INCOLLECTION{Bally_2008,
       author = {{Bally}, J.},
        title = "{Overview of the Orion Complex}",
    booktitle = {Handbook of Star Forming Regions, Volume I},
         year = 2008,
       editor = {{Reipurth}, B.},
       volume = {4},
        pages = {459},
          doi = {10.48550/arXiv.0812.0046},
       adsurl = {https://ui.adsabs.harvard.edu/abs/2008hsf1.book..459B}
}

@ARTICLE{Carpenter_1998,
       author = {{Carpenter}, John M. and {Sanders}, D.~B.},
        title = "{The W51 Giant Molecular Cloud}",
      journal = {\aj},
         year = 1998,
        month = oct,
       volume = {116},
       number = {4},
        pages = {1856-1867},
          doi = {10.1086/300534},
archivePrefix = {arXiv},
       eprint = {astro-ph/9806298},
 primaryClass = {astro-ph},
       adsurl = {https://ui.adsabs.harvard.edu/abs/1998AJ....116.1856C}
}

@ARTICLE{Fujita_2021,
       author = {{Fujita}, Shinji and {Torii}, Kazufumi and {Kuno}, Nario and {Nishimura}, Atsushi and {Umemoto}, Tomofumi and {Minamidani}, Tetsuhiro and {Kohno}, Mikito and {Yamagishi}, Mitsuyoshi and {Tosaki}, Tomoka and {Matsuo}, Mitsuhiro and {Tsuda}, Yuya and {Enokiya}, Rei and {Tachihara}, Kengo and {Ohama}, Akio and {Sano}, Hidetoshi and {Okawa}, Kazuki and {Hayashi}, Katsuhiro and {Yoshiike}, Satoshi and {Tsutsumi}, Daichi and {Fukui}, Yasuo},
        title = "{Massive star formation in W51 A triggered by cloud-cloud collisions}",
      journal = {\pasj},
         year = 2021,
        month = jan,
       volume = {73},
        pages = {S172-S200},
          doi = {10.1093/pasj/psz028},
       adsurl = {https://ui.adsabs.harvard.edu/abs/2021PASJ...73S.172F}
}

@ARTICLE{Wilson_2005,
       author = {{Wilson}, B.~A. and {Dame}, T.~M. and {Masheder}, M.~R.~W. and {Thaddeus}, P.},
        title = "{A uniform CO survey of the molecular clouds in Orion and Monoceros}",
      journal = {\aap},
         year = 2005,
        month = feb,
       volume = {430},
        pages = {523-539},
          doi = {10.1051/0004-6361:20035943},
archivePrefix = {arXiv},
       eprint = {astro-ph/0411089},
 primaryClass = {astro-ph},
       adsurl = {https://ui.adsabs.harvard.edu/abs/2005A&A...430..523W}
}

@inproceedings{Chao-te_2018,
author = {Chao-Te Li and C. M. Bradford and Abigail Crites and Jonathon Hunacek and Tashun Wei and Jen-Chieh Cheng and Tzu-Ching Chang and James Bock},
title = {{TIME millimeter wave grating spectrometer}},
volume = {10708},
booktitle = {Millimeter, Submillimeter, and Far-Infrared Detectors and Instrumentation for Astronomy IX},
editor = {Jonas Zmuidzinas and Jian-Rong Gao},
organization = {International Society for Optics and Photonics},
publisher = {SPIE},
pages = {107083F},
year = {2018},
doi = {10.1117/12.2311415},
URL = {https://doi.org/10.1117/12.2311415}
}

@ARTICLE{Hunacek_2018,
       author = {{Hunacek}, J. and {Bock}, J. and {Bradford}, C.~M. and {Butler}, V. and {Chang}, T.-C. and {Cheng}, Y.-T. and {Cooray}, A. and {Crites}, A. and {Frez}, C. and {Hailey-Dunsheath}, S. and {Hoscheit}, B. and {Kim}, D.~W. and {Li}, C.-T. and {Marrone}, D. and {Moncelsi}, L. and {Shirokoff}, E. and {Steinbach}, B. and {Sun}, G. and {Trumper}, I. and {Turner}, A. and {Uzgil}, B. and {Weber}, A. and {Zemcov}, M.},
        title = "{Hafnium Films and Magnetic Shielding for TIME, A mm-Wavelength Spectrometer Array}",
      journal = {Journal of Low Temperature Physics},
         year = 2018,
        month = dec,
       volume = {193},
       number = {5-6},
        pages = {893-900},
          doi = {10.1007/s10909-018-1906-3},
       adsurl = {https://ui.adsabs.harvard.edu/abs/2018JLTP..193..893H}
}

@PHDTHESIS{Jon_PHD,
       author = {{Hunacek}, Jonathon},
        title = "{Time: A Millimeter-Wavelength Grating Spectrometer Array for [CII] / CO Intensity Mapping}",
       school = {California Institute of Technology, Division of Physics, Mathematics and
        Astronomy},
         year = 2020,
        month = jan,
       adsurl = {https://ui.adsabs.harvard.edu/abs/2020PhDT........69H}
}

@ARTICLE{Sayers_2010,
       author = {{Sayers}, J. and {Golwala}, S.~R. and {Ade}, P.~A.~R. and {Aguirre}, J.~E. and {Bock}, J.~J. and {Edgington}, S.~F. and {Glenn}, J. and {Goldin}, A. and {Haig}, D. and {Lange}, A.~E. and {Laurent}, G.~T. and {Mauskopf}, P.~D. and {Nguyen}, H.~T. and {Rossinot}, P. and {Schlaerth}, J.},
        title = "{Studies of Millimeter-wave Atmospheric Noise above Mauna Kea}",
      journal = {\apj},
         year = 2010,
        month = jan,
       volume = {708},
       number = {2},
        pages = {1674-1691},
          doi = {10.1088/0004-637X/708/2/1674},
archivePrefix = {arXiv},
       eprint = {0904.3943},
 primaryClass = {astro-ph.IM},
       adsurl = {https://ui.adsabs.harvard.edu/abs/2010ApJ...708.1674S}
}

@ARTICLE{lim_2019,
       author = {{Lim}, Wanggi and {De Buizer}, James M.},
        title = "{Surveying the Giant H II Regions of the Milky Way with SOFIA. I. W51A}",
      journal = {\apj},
         year = 2019,
        month = mar,
       volume = {873},
       number = {1},
          eid = {51},
        pages = {51},
          doi = {10.3847/1538-4357/ab0288},
archivePrefix = {arXiv},
       eprint = {1901.07561},
 primaryClass = {astro-ph.GA},
       adsurl = {https://ui.adsabs.harvard.edu/abs/2019ApJ...873...51L}
}

@ARTICLE{Yang2026,
       author = {{Yang}, Selina F. and {McAtee}, Sophie M. and {Vaughan}, Benjamin J. and {Crites}, Abigail T. and {Butler}, Victoria L. and {Chung}, Dongwoo T. and {Keenan}, Ryan P. and {Pham}, Dang and {Prakash}, Shwetha and {Bock}, James J. and {Bradford}, Charles M. and {Chang}, Tzu-Ching and {Cheng}, Yun-Ting and {Dunn}, Audrey and {Emerson}, Nicholas and {Frez}, Clifford and {Hunacek}, Jonathon and {Li}, Chao-Te and {Lowe}, Ian N. and {Lau}, King and {Marrone}, Daniel P. and {Mayer}, Evan C. and {Sun}, Guochao and {Trumper}, Isaac and {Turner}, Anthony D. and {Wei}, Ta-Shun and {Zemcov}, Michael and {TIME Collaboration}},
        title = "{TIME Commissioning Observations. I. Mapping Dust and Molecular Gas in the Sgr A Molecular Cloud Complex at the Galactic Center}",
      journal = {\apj},
         year = 2026,
        month = may,
       volume = {1003},
       number = {1},
          eid = {23},
        pages = {23},
          doi = {10.3847/1538-4357/ae606c},
archivePrefix = {arXiv},
       eprint = {2511.09473},
 primaryClass = {astro-ph.IM},
       adsurl = {https://ui.adsabs.harvard.edu/abs/2026ApJ..1003...23Y}
}

@article{Trumper_2019,
author = {Isaac L. Trumper and Daniel P. Marrone and Dae Wook Kim},
title = {{Utilizing freeform optics in dynamic optical configuration designs}},
volume = {5},
journal = {Journal of Astronomical Telescopes, Instruments, and Systems},
number = {3},
publisher = {SPIE},
pages = {035005},
year = {2019},
doi = {10.1117/1.JATIS.5.3.035005},
URL = {https://doi.org/10.1117/1.JATIS.5.3.035005}
}

@ARTICLE{Ginsburg2013,
       author = {{Ginsburg}, Adam and {Glenn}, Jason and {Rosolowsky}, Erik and {Ellsworth-Bowers}, Timothy P. and {Battersby}, Cara and {Dunham}, Miranda and {Merello}, Manuel and {Shirley}, Yancy and {Bally}, John and {Evans}, II, Neal J. and {Stringfellow}, Guy and {Aguirre}, James},
        title = "{The Bolocam Galactic Plane Survey. IX. Data Release 2 and Outer Galaxy Extension}",
      journal = {\apjs},
         year = 2013,
        month = oct,
       volume = {208},
       number = {2},
          eid = {14},
        pages = {14},
          doi = {10.1088/0067-0049/208/2/14},
archivePrefix = {arXiv},
       eprint = {1305.6622},
 primaryClass = {astro-ph.GA},
       adsurl = {https://ui.adsabs.harvard.edu/abs/2013ApJS..208...14G}
}
\bibliographystyle{aasjournalv7}

\end{document}